\documentclass[
twocolumn
]{aastex631}
\submitjournal{ApJS}

\usepackage{CJKutf8}
\usepackage{hyperref}
\usepackage{amsmath} 
\usepackage{enumerate}
\usepackage{bm}

\shorttitle{resistive GRMHD in \texttt{Gmunu}}
\shortauthors{Cheong et al.}
\graphicspath{{./}{figures/}}

\begin{document}

\title{An extension of \texttt{Gmunu}: General-relativistic resistive magnetohydrodynamics based on staggered-meshed constrained transport with elliptic cleaning}

\author[0000-0003-1449-3363]{Patrick Chi-Kit Cheong \begin{CJK*}{UTF8}{bkai}(張志杰)\end{CJK*}}
\affiliation{Department of Physics, The Chinese University of Hong Kong, Shatin, N.T., Hong Kong}

\author{David Yat Tung Pong}
\affiliation{Department of Physics, The Chinese University of Hong Kong, Shatin, N.T., Hong Kong}

\author{Anson Ka Long Yip}
\affiliation{Department of Physics, The Chinese University of Hong Kong, Shatin, N.T., Hong Kong}

\author[0000-0003-4297-7365]{Tjonnie Guang Feng Li}
\affiliation{Department of Physics, The Chinese University of Hong Kong, Shatin, N.T., Hong Kong}
\affiliation{Institute for Theoretical Physics, KU Leuven, Celestijnenlaan 200D, B-3001 Leuven, Belgium}
\affiliation{Department of Electrical Engineering (ESAT), KU Leuven, Kasteelpark Arenberg 10, B-3001 Leuven, Belgium }



\begin{abstract}
	We present the implementation of general-relativistic \emph{resistive} magnetohydrodynamics solvers and three divergence-free handling approaches adopted in the \texttt{G}eneral-relativistic \texttt{mu}ltigrid \texttt{nu}merical (\texttt{Gmunu}) code.
	In particular, implicit-explicit Runge-Kutta schemes are used to deal with the stiff terms in the evolution equations for small resistivity.
	The three divergence-free handling methods are (i) hyperbolic divergence cleaning (also known as the generalised Lagrange multiplier); (ii) staggered-meshed constrained transport schemes and (iii) elliptic cleaning through a multigrid solver which is applicable in both cell-centred and face-centred (stagger grid) magnetic fields.
	The implementation has been tested with a number of numerical benchmarks from special-relativistic to general-relativistic cases.
	We demonstrate that our code can robustly recover from ideal magnetohydrodynamics limit to a highly resistive limit.
	We also illustrate the applications in modelling magnetised neutron stars, and compare how different divergence-free handling affects the evolution of the stars.
	Furthermore, we show that the preservation of the divergence-free condition of the magnetic field when using staggered-meshed constrained transport schemes can be significantly improved by applying elliptic cleaning.
\end{abstract}



\section{\label{sec:intro}Introduction}
Accurate and realistic modelling of the dynamics of relativistic plasmas is extremely important for understanding high energy astrophysical phenomena.
Examples of these relativistic astrophysical systems are pulsars, magnetars, gamma-ray bursts and active galactic nuclei.
Typically, since the Ohmic diffusion timescale of the magnetised plasma in most of the astrophysical scenarios is much longer then the characteristic dynamical timescale the large-scale dynamics of these plasmas can be well described by neglecting the dissipative processes, namely, in the ideal magnetohydrodynamics (MHD) limit.
Progress on the development of robust and accurate numerical codes for solving ideal relativistic MHD systems have been achieved over the last decade, e.g., \cite{1999MNRAS.303..343K, 2001ApJS..132...83B, 2002A&A...390.1177D, 2003ApJ...589..444G, 2006MNRAS.368.1040M, 2006JFM...562..223G, 2007A&A...473...11D, 2009MNRAS.393.1141M, 2017ComAC...4....1P, 2019A&A...629A..61O, 2019ApJS..244...10R, 2019arXiv191210192L, 2020PhRvD.101j4007M, 2020CQGra..37m5010C}.
In addition, these relativistic MHD codes have been successfully applied in studies of gamma-ray bursts and relativistic jets \cite{2009MNRAS.394L.126M,2009A&A...494..879M,2010MNRAS.402....7M, 2015ApJ...809...38M,2016MNRAS.462.3031B,2017A&A...606A..57R, 2018MNRAS.475.2971B,2019ApJ...870L..20N}, accretion flows \cite{2012MNRAS.423.3083M, 2013MNRAS.435..718M, 2018NatAs...2..585M} and magnetised relativistic stars \cite{2021arXiv210811858S, 2007PhRvD..76b4019Z,2012arXiv1203.3590L, 2013PhRvD..88j3005L}.

Although the ideal MHD approximation has been extensively used to model astrophysical systems, going beyond this approach is required for more accurate and realistic modelling of plasmas.
In particular, assuming vanishing electrical resistivity prevents some important physical processes such as dissipation and magnetic reconnection.
Magnetic reconnection can change the magnetic field's topology, and convert magnetic energy into other forms of energy such as heat and kinetic energy. 
These processes, although usually occurring at very small length-scales, could significantly affect the large scale dynamics of the plasmas.

One natural approach to overcome these limitations is to consider the resistive MHD framework.
In this framework, the full Maxwell and hydrodynamic equations are solved.
The coupling between electromagnetic fields and fluid are determined by the electric current, typically provided by Ohm's law.
A suitable choice of current enables us to describe physical dissipation and reconnection at different region by resolving plasmas at both the ideal MHD limit (the resistivity $\eta \rightarrow 0$) and the finite resistive regime.

Numerically solving resistive MHD is more challenging than ideal MHD due to the appearance of the stiff relaxation term.
Unlike in the ideal MHD approximation, where the electric field $E^i$ is purely dependent variable (i.e., $E^i = - \epsilon^{ijk} v_j B_k$), the electric field $E^i$ is an independent variable in resistive MHD framework.
When the resistivity is realistically small but finite, the dynamical time-scales of the electric field is much shorter than the MHD evolution, resulting in a stiff relaxation term in Amp{\'e}re's law.
Applying explicit time integration would be inefficient due to the extremely strict constraints on the time steps due to the stiff relaxation terms.
Implicit-explicit (IMEX) Runge-Kutta schemes (see, e.g., \cite{pareschi2005implicit}) offer an effective approach to overcome this problem.
These schemes have been applied and tested in several codes \cite{2009MNRAS.394.1727P,2013MNRAS.428...71B,2013PhRvD..88d4020D,2017ApJ...834...29Q,2018MNRAS.476.3837M,2019MNRAS.486.4252M,2019ApJS..244...10R}.

In this work, we extend \texttt{Gmunu} by implementing the IMEX based general relativistic resistive MHD framework.
In addition to elliptical divergence cleaning for magnetic field, which removes the magnetic monopoles by solving Poisson's equation, established in our previous work \cite{2021MNRAS.tmp.2378C}, we included two more approaches to handle the solenoidal condition, namely, (i) hyperbolic divergence cleaning through a generalised Lagrange multiplier (GLM) (see a Newtonian example in \cite{DEDNER2002645}); (ii) constrained transport (CT) scheme which updates the magnetic fields while controlling the divergence-free constraint to numerical round-off accuracy \cite{1988ApJ...332..659E}.
Finally, our code is validated through some benchmarking tests.

The paper is organised as follows.
In section~\ref{sec:formulations} we outline the formalism we used in this work, including the details of the numerical settings, and the methodology, and implementation of our resistive magnetohydrodynamics solver.
The implementation of three divergence-free handling approaches are presented in section~\ref{sec:divb}.
The code tests and results are presented in section~\ref{sec:numerical_tests}.
This paper ends with a discussion in section~\ref{sec:conclusions}.

{Unless} explicitly stated, we work in geometrized {Heaviside}-Lorentz units, {for} which the speed of light $c$, gravitational constant $G$, solar mass $M_{\odot}$, vacuum permittivity $\epsilon_0$ and vacuum permeability $\mu_0$ are all {equal} to one ( $c=G=M_{\odot}=\epsilon_0 = \mu_0 = 1$ ).
Greek indices, running from 0 to 3, are used for 4-quantities while the Roman indices, running from 1 to 3, are used for 3-quantities.

\section{Formulation and Numerical methods}\label{sec:formulations} 
\subsection{Resistive GRMHD in the reference-metric formalism}\label{sec:GREMHD_equations}
We use the standard Arnowitt-Deser-Misner (ADM) 3+1 formalism \citet{2007gr.qc.....3035G,2008itnr.book.....A}. 
The metric can be written as
\begin{equation}
	\begin{aligned}
	ds^2 = & g_{\mu\nu}dx^\mu dx^\nu \\
		= & -\alpha^2  dt^2 + \gamma_{ij} \left( dx^i + \beta^i dt \right) \left( dx^j + \beta^j dt \right)
	\end{aligned}
\end{equation}
where $\alpha$ is the lapse function, $\beta^i$ is the spacelike shift vector and $\gamma_{ij}$ is the spatial metric.
We adopt a conformal decomposition of the spatial metric $\gamma_{ij}$ with the conformal factor $\psi$:
\begin{equation}
	\gamma_{ij} = \psi^4 \bar{\gamma}_{ij},
\end{equation}
where $\bar{\gamma}_{ij}$ is the conformally related metric.

The evolution equations for matter are derived from the local conservation of the rest-mass and energy-momentum, and the Maxwell equations:
\begin{align}
	&\nabla_\mu \left( \rho u^\mu \right) = 0, \\
	&\nabla_\mu T^{\mu\nu} = 0, \\
	&\nabla_\mu F^{\mu\nu} = - \mathcal{J}^\nu \label{eq:maxwell_eq1},\\
	&\nabla_\mu {}^*F^{\mu\nu} = 0  \label{eq:maxwell_eq2},
\end{align}
where $\rho$ is the rest-mass density of the fluid, $u^\mu$ is the fluid four-velocity, $F^{\mu\nu}$ is the Maxwell tensor, ${}^*F^{\mu\nu}$ is the Faraday tensor.
$T^{\mu\nu}$ is the total energy-momentum tensor, which is the sum of the energy-momentum tensor of fluid and electromagnetic field:
\begin{equation}
	T^{\mu\nu} = T^{\mu\nu}_{\text{fluid}} + T^{\mu\nu}_{\text{EM}}.
\end{equation}
Here, the energy-momentum tensor of fluid is
\begin{equation}
	T^{\mu\nu}_{\text{fluid}} = \rho h u^{\mu} u^{\nu} + p g^{\mu\nu},
\end{equation}
where  $ h \equiv 1 + \epsilon + p / \rho$ is the specific enthalpy, $p$ is pressure and $\epsilon$ is the specific energy, while the energy-momentum tensor of the electric and magnetic field $T^{\mu\nu}_{EM}$ is
\begin{equation}
	\begin{aligned}
		T^{\mu\nu}_{\text{EM}} = &F^{\mu\alpha} F^{\mu}_{\;\alpha} - \frac{1}{4}g^{\mu\nu} F_{\alpha\beta}F^{\alpha\beta}. \label{eq:Tmunu_EM}
	\end{aligned}
\end{equation}

The electric and magnetic fields are defined as:
\begin{align}
	E^{\mu} := F^{\mu\nu} n_\nu , \quad B^{\mu} := {}^*F^{\mu\nu} n_\nu,
\end{align}
where $n^\mu$ is the 4-velocities of an Eulerian observer.
Note that both electric and magnetic field are purely spatial, i.e., $E^{\mu}n_\mu = 0 $ and $B^{\mu}n_\mu = 0$.
It is useful to introduce the electric and magnetic field in fluid-frame (comoving-frame) 
\begin{align}
	&e^\mu := F^{\mu\nu} u_\nu = W n^\mu \left(E^iv_i\right) + W\left(E^\mu + \epsilon^{\mu\nu\lambda} v_\nu B_\lambda\right), \label{eq:E_comoving} \\
	&b^\mu := {}^*F^{\mu\nu} u_\nu = W n^\mu \left(B^iv_i\right) + W\left(B^\mu - \epsilon^{\mu\nu\lambda} v_\nu E_\lambda\right), \label{eq:B_comoving}
\end{align}
the energy-momentum tensor $T^{\mu\nu}_{EM}$ can then be written as
\begin{equation}
	\begin{aligned}
		T^{\mu\nu}_{\text{EM}}=&\left(b^2 + e^2\right) \left( u^\mu u^\nu + \frac{1}{2}g^{\mu\nu}\right) - b^\mu b^\nu - e^\mu e^\nu \\
			& - \alpha u_\lambda e_\beta b_\kappa \left( u^\mu \epsilon^{\nu\lambda\beta\kappa} + u^\nu \epsilon^{\mu\lambda\beta\kappa} \right),
	\end{aligned}
\end{equation}
where $e^2 := e^\mu e_\mu$ and $b^2 := b^\mu b_\mu$.

Under $3+1$ decomposition, the electric 4-current $\mathcal{J}^\nu$ can be expressed as
\begin{equation}
	\mathcal{J}^\mu = n^\mu q + J^\mu,
\end{equation}
where $ \rho_e = -\mathcal{J}^\mu n_\mu$ is the charge density, $J^\mu$ is the purely spatial (i.e., $J^\mu n_\mu = 0$) current density observed by an Eulerian observer moving with 4-velocities $n^\mu$.

As in \cite{2021MNRAS.tmp.2378C}, under the reference-metric formalism, the evolution equations can be expressed as:
\begin{equation} \label{eq:GRMHD_ref}
	\partial_t \bm{q} + \frac{1}{\sqrt{\hat{\gamma}}}\partial_j\left[\sqrt{\hat{\gamma}} \bm{f}^j\right] = \bm{s} + \bm{s}_{\text{geom}} ,
\end{equation}
where $\bm{s}_{\text{geom}}$ are so-called geometrical source terms which contain the 3-Christoffel symbols $ \hat{\Gamma}^l_{ik} $ associated with the reference metric $\hat{\gamma}_{ij}$, and
\begin{equation}
	\bm{q} = \begin{bmatrix}
           q_D \\
           q_{S_j} \\
           q_{\tau} \\
           q_{{B}^j} \\
           q_{{E}^j}
        \end{bmatrix}, 
	\bm{f^i} = 
        \begin{bmatrix}
           \left(f_D\right)^i \\
           \left(f_{S_j}\right)^i \\
           \left(f_\tau\right)^i \\
           \left(f_{B^j}\right)^i \\
           \left(f_{E^j}\right)^i \\
        \end{bmatrix} ,
	\bm{s} = 
        \begin{bmatrix}
           s_D \\
           s_{S_j} \\
           s_{\tau} \\
           s_{{B}^j} \\
           s_{{E}^j}
        \end{bmatrix}
	.
\end{equation}
Here, $\bm{q}$ are the conserved quantities:
\begin{align}
	q_D :=& \psi^6 \sqrt{\bar{\gamma}/\hat{\gamma}} D =  \psi^6 \sqrt{\bar{\gamma}/\hat{\gamma}} \left[ \rho W  \right] ,\\
	q_{S_j} :=&  \psi^6 \sqrt{\bar{\gamma}/\hat{\gamma}} S_j \\
		= & \psi^6 \sqrt{\bar{\gamma}/\hat{\gamma}} \left[  \rho h W^2 v_j + \sqrt{{\gamma}}\eta_{jkl} E^k B^l \right] , \nonumber \\
	q_{\tau} :=&  \psi^6 \sqrt{\bar{\gamma}/\hat{\gamma}} \tau \\
		= &  \psi^6 \sqrt{\bar{\gamma}/\hat{\gamma}} \left[  \rho h W^2 - p + \frac{1}{2}\left( E^2 + B^2 \right) - D \right] , \nonumber \\
	q_{{B}^j} :=& \psi^6 \sqrt{\bar{\gamma}/\hat{\gamma}} B^j  , \\
	q_{{E}^j} :=& \psi^6 \sqrt{\bar{\gamma}/\hat{\gamma}} E^j  , 
\end{align}
where $\eta^{ijk} = \eta_{ijk}$ is the 3-dimensional Levi-Civita symbol.
The corresponding fluxes $\bm{f}^i$ are given by:
\begin{align}
           \left(f_D\right)^i  :=& \psi^6 \sqrt{\bar{\gamma}/\hat{\gamma}} \left[  D \hat{v}^i \right] \\
           \left(f_{S_j}\right)^i  :=& \psi^6 \sqrt{\bar{\gamma}/\hat{\gamma}} \left[ - S_j \beta^i + \alpha S^i_{\;j} \right] \\
            = & \psi^6 \sqrt{\bar{\gamma}/\hat{\gamma}} \Big[ - S_j \beta^i + \alpha \Big( \left[ p + \frac{1}{2}\left(E^2 + B^2\right)\right]\delta^i_j \nonumber \\ 
		& + \rho h W^2 v^i v_j - E^i E_j - B^i B_j \Big) \Big] ,  \nonumber \\
           \left(f_\tau\right)^i  :=& \psi^6 \sqrt{\bar{\gamma}/\hat{\gamma}}  \left[ - \tau \beta^i + \alpha \left( S^i - v^i D  \right) \right] , \\
           \left(f_{B^j}\right)^i  := & \psi^6 \sqrt{\bar{\gamma}/\hat{\gamma}}  \left[ \epsilon^{jik} \hat{E}_k  \right] \\
	= & \psi^6 \sqrt{\bar{\gamma}/\hat{\gamma}}  \left[ \beta^j {B}^i -  \beta^i {B}^j - \frac{1}{\sqrt{{\gamma}}} \eta^{ijk} \alpha E_k  \right], \\ \nonumber
           \left(f_{E^j}\right)^i  :=& \psi^6 \sqrt{\bar{\gamma}/\hat{\gamma}}  \left[ - \epsilon^{jik} \hat{B}_k  \right] \\
		= &\psi^6 \sqrt{\bar{\gamma}/\hat{\gamma}}  \left[ \beta^j {E}^i -  \beta^i {E}^j + \frac{1}{\sqrt{{\gamma}}} \eta^{ijk} \alpha B_k  \right], \nonumber 
\end{align}
where we have defined 
\begin{align}
	\hat{E}_i :=& \alpha E_i + \epsilon_{ijk}\beta^j B^k \label{eq:e_hat}, \\
	\hat{B}_i :=& \alpha B_i - \epsilon_{ijk}\beta^j E^k \label{eq:b_hat},
\end{align}
and $\epsilon^{ijk}$ is the spacial Levi-Civita pseudo-tensor, which can be expressed with the 3-dimensional Levi-Civita symbol $\eta^{ijk}$ by
\begin{equation}
	\epsilon^{ijk} = \frac{1}{\sqrt{\gamma}}\eta^{ijk}, \qquad
	\epsilon_{ijk} = {\sqrt{\gamma}}\eta_{ijk}.
\end{equation}
Finally, the corresponding source terms $\bm{s}$ are given by:
\begin{align}
	&s_D = 0 , \\
	&\begin{aligned}
        s_{S_i} = &\alpha \psi^6 \sqrt{\bar{\gamma}/\hat{\gamma}} \Big\{ 
                  - T^{00} \alpha \partial_i \alpha + T^{0}_k \hat{\nabla}_i \beta^k \\ 
                  & + \frac{1}{2} \left( T^{00} \beta^j \beta^k + 2 T^{0j}\beta^k + T^{jk} \right) \hat{\nabla}_i \gamma_{jk} \Big\} , 
	\end{aligned} \\
	&\begin{aligned}
        s_\tau = &\alpha \psi^6 \sqrt{\bar{\gamma}/\hat{\gamma}} \Big\{ T^{00} \left( K_{ij} \beta^i \beta^j - \beta^k \partial_k \alpha \right) \\
        	&+ T^{0j} \left( 2 K_{jk} \beta^k - \partial_j \alpha \right) + T^{ij}K_{ij} \Big\} ,
	\end{aligned} \\
	&s_{B^i} = 0 , \\
	&s_{E^i} = \psi^6 \sqrt{\bar{\gamma}/\hat{\gamma}} \left\{ \rho_e \beta^i - \alpha J^i \right\} \label{eq:amp_source} ,  
\end{align}
where $K_{ij}$ is the extrinsic curvature, $ \rho_e $ is the electric charge density, $J^j$ is the 3-current.

In addition to the evolution equations, the Maxwell equations~\eqref{eq:maxwell_eq1} and \eqref{eq:maxwell_eq2} give two constrain equations, namely, Gauss's law and the solenoidal constraint equation
\begin{align}
	&\hat{\nabla}_i q_{E^i} = \frac{1}{\sqrt{\hat{\gamma}}} \partial_i \left( \sqrt{\hat{\gamma}} q_{E^i} \right) = \psi^6 \sqrt{\bar{\gamma}/\hat{\gamma}} \rho_e, \label{eq:gauss_law} \\
	&\hat{\nabla}_i q_{B^i} = \frac{1}{\sqrt{\hat{\gamma}}} \partial_i \left( \sqrt{\hat{\gamma}} q_{B^i} \right) = 0. \label{eq:divb_0}
\end{align}

Note that, to completely close Maxwell's equations, we need to determine the electric charge density $\rho_e$ and the 3-current $J^j$ which describes the coupling between the electromagnetic fields and the fluid.
In practice, the charge density $\rho_e$ is determined by solving the continuity equation of the charge (i.e., $\nabla_\mu \mathcal{J}^\mu = 0$) or by Gauss's law~\eqref{eq:gauss_law}.
In the current implementation, we use the later approach, namely, substituting equation~\eqref{eq:gauss_law} into the source term of the Amp{\'e}re's law (equation~\eqref{eq:amp_source}).
On the other hand, the 3-current $J^j$ is determined by Ohm's law, which is discussed in section~\ref{sec:ohms_law}.

\subsection{\label{sec:ohms_law}Generalised covariant Ohm’s law}
Usually, Ohm's law for isotropic resistive general-relativistic magneto-hydrodynamics can be expressed as a linear relation between the comoving electric field $e^\mu$ and current density $j^\mu$
\begin{equation}
	e^\mu = \eta j^\mu,
\end{equation}
where $\eta$ the electrical resistivity.
It is also useful to define the electrical conductivity of the medium $\sigma_\textrm{c}$ as the inverse of the resistivity, i.e., $\sigma_\textrm{c} := \eta^{-1}$.
For an ideal plasma with infinite conductivity $\eta = 0$, the comoving electric field vanishes (i.e., $e^\mu = 0$) which is the ideal general-relativistic magneto-hydrodynamics.
Additionally, a mean-field $\alpha$-dynamo effect can be included by generalizing this Ohm's law by \cite{2013MNRAS.428...71B}
\begin{equation} \label{eq:comoving_ohms_law}
	e^\mu = \xi b^\mu + \eta j^\mu,
\end{equation}
where $\xi$ is the $\alpha$-dynamo parameter, here we follow the notation in \cite{2013MNRAS.428...71B} to avoid the conflict with the lapse function $\alpha$.
Ohm's law~\eqref{eq:comoving_ohms_law} can be written in 3+1 form as
\begin{equation}\label{eq:ohms_law}
	\begin{aligned}
		J^i = \rho_e v^i + \frac{W}{\eta} \Bigg\{ & \left[ E^i + \epsilon^{ijk}v_jB_k - \left( E^j v_j \right) v^i\right] \\
		& - \xi \left[ B^i - \epsilon^{ijk}v_jE_k - \left( B^j v_j \right) v^i \right] \Bigg\} .
	\end{aligned}
\end{equation}
Amp{\'e}re's law for the electric field evolution can now be closed by substituting Ohm's law~\eqref{eq:ohms_law} into the source term of it (equation~\eqref{eq:amp_source}).

\subsection{\label{sec:imex}Time stepping: Implicit-explicit Runge-Kutta schemes}
The relativistic resistive magnetohydrodynamics system is a stiff system, i.e., hyperbolic systems with relaxation terms.
In particular, the source term of the modified Amp{\'e}re's law mentioned here becomes stiff when the conductivity $\sigma_\textrm{c}$ is very large yet finite (or when the resistivity $\eta$ is very small but finite, see equation~\ref{eq:ohms_law} in section~\ref{sec:ohms_law}).
If an explicit time integration method is applied (which is widely used in ideal general relativistic magnetohydrodynamics), the time-steps would be extremely small because it scales with the resistivity, thus it requires high computational costs.
One possible method to tackle this is to solve the stiff part \emph{implicitly} without further limiting the time-step sizes.
The implicit-explicit Runge-Kutta (IMEX) methods, which are shown to be effective and robust for relativistic systems with stiff terms contained, will be outlined below by following \cite{pareschi2005implicit}.

Consider a stiff system with the form
\begin{equation}
	\partial_t {\bm{q}} = \mathcal{\bm{L}}(\bm{q}) + \frac{1}{\epsilon}\mathcal{\bm{R}}(\bm{q}),
\end{equation}
where $\mathcal{\bm{L}}(\bm{q})$ is representing the non-stiff part of the conservation equations, including the advection terms and the non-stiff source terms, while $\mathcal{\bm{R}}(\bm{q}) / \epsilon$ is the \emph{stiff} term with a \emph{relaxation parameter} $\epsilon$.
An IMEX Runge-Kutta scheme can be expressed as
\begin{align}
	\left<\bm{q}\right>^{(i)} = \left<\bm{q}\right>^{\texttt{n}} 
	& - \Delta t \sum_{j=0}^{i-1}\left[ \tilde{a}_{ij}\mathcal{\bm{L}}\left(\left<\bm{q}\right>^{(j)}\right)\right]  \nonumber \\
	& + \Delta t \sum_{j=0}^{N_{\text{IMEX}}}\left[ {a}_{ij}\frac{1}{\epsilon}\mathcal{\bm{R}}\left(\left<\bm{q}\right>^{(j)}\right)\right], \\
	\left<\bm{q}\right>^{\texttt{n+1}} = \left<\bm{q}\right>^{\texttt{n}} 
	& - \Delta t \sum_{j=0}^{N_{\text{IMEX}}}\left[ \tilde{w}_{i}\mathcal{\bm{L}}\left(\left<\bm{q}\right>^{(i)}\right)\right]  \nonumber \\
	& + \Delta t \sum_{j=0}^{N_{\text{IMEX}}}\left[ {w}_{i}\frac{1}{\epsilon}\mathcal{\bm{R}}\left(\left<\bm{q}\right>^{(i)}\right)\right],
\end{align}
with matrices $\tilde{A} = \left(\tilde{a}_{ij}\right)$ where $\tilde{a}_{ij}=0$ for $j \leq i$ and ${A} = \left({a}_{ij}\right)$ are $N_{\text{IMEX}} \times N_{\text{IMEX}}$ matrices such that the whole scheme is \emph{explicit} in $\mathcal{\bm{L}}$ and implicit in $\mathcal{\bm{R}}$. 
In addition, $\tilde{w}_i$ and $w_i$ are some constant coefficients. 
Alternatively, the IMEX Runge-Kutta schemes can be represented in terms of Butcher notation,
\begin{center}
	\begin{tabular}{ c | c }
		\multicolumn2c{Explicit} \\
		$\tilde{c}_i$ & $\tilde{a}_{ij}$  \\ \hline
		  & $\tilde{w_i}^T$ 
	\end{tabular}
	\quad
	\begin{tabular}{ c | c }
		\multicolumn2c{Implicit} \\
		${c}_i$ & ${a}_{ij}$  \\ \hline
		  & ${w_i}^T$ 
	\end{tabular}
	\quad, 
\end{center}
where ${w_i}^T$ denote the transposition of ${w_i}$; $\tilde{c}_i := \sum\limits^{i-1}_{j=1}\tilde{a}_{ij}$ and $c_i := \sum\limits^{i-1}_{j=1} a_{ij}$, which are not used in the actual implementation.
Obviously, the matrices $\tilde{A}$ and ${A}$ (and so as $\tilde{c}_i$ and $c_i$ ) together with $\tilde{w}_i$ and $w_i$ need to be chosen to satisfy the order conditions.

The \emph{explicit} part of the IMEX Ruuge-Kutta schemes can be constructed to be strongly stability preserving (SSP).
We shall use the notation SSP$k\left( s, \sigma, p \right)$ to denote different schemes, where $k$ represent the order of the SSP scheme, $s$ denote the number of stages of the \emph{implicit} scheme, $\sigma$ denotes the stages of the \emph{explicit} scheme and finally $p$ denote the order of the IMEX scheme.
For example, tables~\ref{tab:imex_ssp2} show the IMEX-SSP2(2,2,2) scheme \cite{pareschi2005implicit}.
\begin{table}[h]
	\centering
	\caption{Tableau for the IMEX-SSP2(2,2,2) L-Stable Scheme \cite{pareschi2005implicit}.}
	\label{tab:imex_ssp2}
	\begin{tabular}{ c c }
		\begin{tabular}{ c | c c }
			\multicolumn{3}{c}{Explicit} \\
			0 & 0 & 0 \\
			1 & 1 & 0 \\ \hline
			  & 1/2 & 1/2 \\ 
		\end{tabular}
		& 
		\begin{tabular}{ c | c c }
			\multicolumn{3}{c}{Implicit} \\
			$\gamma$ & $\gamma$ & 0 \\
			$1-\gamma$ & $1-2\gamma$ & $\gamma$ \\ \hline
			  & 1/2 & 1/2 \\ 
		\end{tabular}
	\end{tabular}
	where $\gamma = 1 - 1/\sqrt{2}$.
\end{table}

\subsection{\label{sec:im_step}Implicit step}
Since only Amp{\'e}re's law (the evolution equation of electric field) contains stiff terms, all the conserved quantities are solved in the explicit steps except for the electric fields.
To apply IMEX schemes, we split the 3-current $J^i$ into two stiff and non-stiff part, namely,
\begin{align}
	&J^i_{\text{non-stiff}} := \rho_e v^i \\
	&J^i_{\text{stiff}} := \frac{W}{\eta} \Bigg\{ \left[ E^i + \epsilon^{ijk}v_jB_k - \left( E^j v_j \right) v^i\right]\\ 
	& - \xi \left[ B^i - \epsilon^{ijk}v_jE_k - \left( B^j v_j \right) v^i \right] \Bigg\}  \nonumber .
\end{align}
The stiff part (the 3-current $J^i_{\text{stiff}}$ term) of the evolution equation of the electric field $E^i$ needed to be solved implicitly, both the updated electric field and 3-velocities are unknown.
In other words, the conversion form conserved to primitive variables and the implicit update of the electric field needs to be solved \emph{simultaneously}.
Several approaches are proposed and explored recently \cite{2019ApJS..244...10R}.
In this work, we adopt the 3D Newton-Raphson approached presented in \cite{2013MNRAS.428...71B, 2020MNRAS.491.2346T}, which has proved to be the most robust approach in various codes \cite{2019ApJS..244...10R, 2019MNRAS.486.4252M}.
Note that this approach may not work for an arbitrary or tabulated equation of state.
Here, we consider the ideal-gas equation of state of perfect fluids $p=\left(\Gamma-1\right)\rho \epsilon$ with polytropic index $\Gamma$, and we define $\hat{\Gamma}_1 := {\Gamma}/\left(\Gamma-1\right)$.

The implicit update for the electric field from time level $n$ to $n+1$ can be expressed as
\begin{equation}
\begin{aligned}
	E^{i,n+1} = & E^{i,*}\\
 + \frac{W}{\tilde{\eta}}\Big\{ & E^{i,n+1} + \epsilon^{ijk}v_j^{n+1} B_k^{n+1} - \left( E^{k,n+1}v_{k}^{n+1} \right) v^{i,n+1} \\ 
	& \xi \Big[ B^{i,n+1} - \epsilon^{ijk} v_j^{n+1} E_k^{n+1}  - \left( B^{k,n+1}v_{k}^{n+1} \right) v^{i,n+1}\Big] \Big\} , 
\end{aligned}
\end{equation}
where $\tilde{\eta} := \eta/ \left(\alpha \Delta t\right)$, and $E^{i,*}$ is the explicit update (with the non-stiff current $J^i_{\text{non-stiff}}$ only) electric field from $E^{i,n}$.
By introducing the normalised 3-velocity $\tilde{u}^i := W v^i$ (so that $W^2 = 1 + \tilde{u}^i\tilde{u}_i$), this equation can be written as (see appendix A in \cite{2020MNRAS.491.2346T})
\begin{equation} \label{eq:im_E}
\begin{aligned}
	A_0 E^{i,n+1} = & \tilde{\eta} E^{i,*} + A_1 \left( E^{k,*}\tilde{u}_{k}^{n+1} \right) \tilde{u}^{i,n+1} + A_2 \epsilon^{ijk} \tilde{u}_j^{n+1} E_k^{*} \\
	&+ A_3 B^{i,n+1} + A_4 \left( B^{k,n+1}\tilde{u}_{k}^{n+1} \right) \tilde{u}^{i,n+1} \\
	&+ A_5 \epsilon^{ijk} \tilde{u}_j^{n+1} B_k^{n+1},
\end{aligned}
\end{equation}
here, six newly defined coefficients are
\begin{align}
	&A_0(W) = W + \tilde{\eta} + \xi^2 \frac{W^2-1}{W+\tilde{\eta}},\\ 
	&A_1(W) = \frac{\tilde{\eta}}{1+\tilde{\eta} W } + \xi^2 \frac{\tilde{\eta}W}{\left(W+\tilde{\eta}\right)\left(1+W\tilde{\eta}\right)},\\ 
	&A_2(W) = -\xi \frac{\tilde{\eta}}{W+\tilde{\eta}},\\ 
	&A_3(W) = \xi \frac{1+\tilde{\eta}W}{W+\tilde{\eta}},\\ 
	&A_4(W) = \xi \frac{1-\tilde{\eta}^2 + \xi^2}{\left(W+\tilde{\eta}\right)\left(1+W\tilde{\eta}\right)},\\ 
	&A_5(W) = -1 - \xi^2 \frac{W}{W+\tilde{\eta}},
\end{align}
and their corresponding derivatives are
\begin{align}
	&\dot{A}_0(W) = 1+ \xi^2 \frac{1+W^2+2W\tilde{\eta}}{\left(W+\tilde{\eta}\right)^2}, \\
	&\dot{A}_1(W) = - \frac{\tilde{\eta}^2}{\left(1+W\tilde{\eta}\right)^2} - \xi^2 \frac{\tilde{\eta}^2\left( W^2-1\right)}{\left(1+W\tilde{\eta}\right)^2\left(W+\tilde{\eta}\right)^2}. \\
	&\dot{A}_2(W) = \xi \frac{\tilde{\eta}}{\left(W+\tilde{\eta}\right)^2}, \\
	&\dot{A}_3(W) = \xi \frac{\tilde{\eta}^2-1}{\left(W+\tilde{\eta}\right)^2}, \\
	&\dot{A}_4(W) = - \xi \frac{\left( 1 - \tilde{\eta}^2 + \xi^2 \right)\left( 1 + \tilde{\eta}^2 + 2W\tilde{\eta} \right)}{\left(1+W\tilde{\eta}\right)^2\left(W+\tilde{\eta}\right)^2}, \\
	&\dot{A}_5(W) = - \xi^2 \frac{\tilde{\eta}}{\left(W+\tilde{\eta}\right)^2}.
\end{align}

The normalised velocity $\tilde{u}^i$ can be obtained by solving 
\begin{equation}
	f_i\left( {\tilde{u}^j}\right) = \tilde{w} \gamma_{ij}\tilde{u}^j - \left( S_i - \epsilon_{ilm}E^lB^m \right), \label{eq:f_u_for_e}
\end{equation}
where the electric field $E^l$ is obtained by evaluating equation~\eqref{eq:im_E}
The modified enthalpy $w$ is
\begin{equation}
	\tilde{w} := \rho h W = \frac{W\left(\tau + D - U_\textrm{EM}\right) - D/\hat{\Gamma}_1}{W^2 - 1/\hat{\Gamma}_1},
\end{equation}
and $U_\textrm{EM}$ is the energy density of the electromagnetic fields $U_\textrm{EM}:= \frac{1}{2}\left(E^2+B^2\right)$.
The Jacobian of $f_i(\tilde{u}^j)$ can be evaluated analytically as
\begin{equation}
	J_{ij} = \frac{\partial f_i}{\partial \tilde{u}^j} = \tilde{w} \gamma_{ij} + \tilde{u}_i \frac{\partial \tilde{w} }{\partial \tilde{u}^j} + \epsilon_{ilm}\frac{\partial E^i}{\partial \tilde{u}^j} B^m,
\end{equation}
where, with the fact that $\partial W / \partial \tilde{u}^j = \tilde{u}^j / W$, one can write
\begin{equation}
	\frac{\partial \tilde{w}}{\partial \tilde{u}^j} = \frac{ \left(\tau + D - U_\textrm{EM}\right)\tilde{u}_j/W - 2\tilde{w}\tilde{u}_j - W\frac{\partial E^i}{\partial \tilde{u}^j} }{W^2 - 1/\hat{\Gamma}_1},
\end{equation}
and
\begin{equation}
	\begin{aligned}
		A_0 W\frac{\partial E^i}{\partial \tilde{u}^j} = 
		& -\dot{A}_0 E^i u_j + A_1 W u^i E_j^* 
		+ \dot{A}_3 B^i u_j + A_4 W u^iB_j \\
		& + \left( A_1 W \gamma^i_j + \dot{A}_1 u^iu_j\right)\left( E^*_ku^k\right) \\
		& + \epsilon^{ilm}\left( A_2 W \gamma_{lj} + \dot{A}_2 u_lu_j \right)E^*_m \\
		& + \left( A_4 W \gamma^i_j + \dot{A}_4 u^iu_j\right)\left( B_ku^k\right) \\
		& + \epsilon^{ilm}\left( A_5 W \gamma_{lj} + \dot{A}_5 u_lu_j \right)B_m . 
	\end{aligned}
\end{equation}

The 3D system on $\tilde{u}^i$ (equation~\eqref{eq:f_u_for_e}) can be solved by multi-dimensional Newton-Raphson method.
For instance, at any iteration $(k)$, the normalised velocity $\tilde{u}^{j\;(k+1)}$ can be obtained by
\begin{equation}
	\tilde{u}^{j\;(k+1)} = \tilde{u}^{j\;(k)} - \left[J_{ij}^{(k)}\right]^{-1} f_i^{(k)},
\end{equation}
and repeat the iterations until the desired tolerance is reached.
Note that Newton-Raphson method occasionally fail to converge if the initial guess is too far away from a root.
To increase its robustness, by following \cite{press1996numerical}, the globally convergent multi-dimensional Newton method is implemented in \texttt{Gmunu}.

\subsection{\label{sec:chara_speed_for_gremhd}Characteristic speed}
The characteristic velocities can be expressed by the following form
\begin{align}
	\lambda^i_{\pm} = \alpha \bar{\lambda}^i_{\pm} - \beta^i,
\end{align}
with the characteristic velocity in the $i$-th direction in the locally flat frame $\alpha \rightarrow 1$, $\beta^j \rightarrow 0$ \cite{anile_1990, 2007A&A...473...11D}.
For simplicity, we assume that the fastest waves locally propagate with speed of light, thus the characteristic to be in the limit of maximum diffusivity \cite{2007A&A...473...11D, 2013MNRAS.428...71B},
\begin{align}
	\bar{\lambda}^i_{\pm} = \pm \sqrt{\gamma^{ii}}.
\end{align}

\subsection{\label{sec:con2prim_gremhd}Conserved to primitive variables conversion}
If the electric field $E^i$ and the magnetic field $B^i$ are given, the conserved variables $\bm{q}$ can then be transformed into primitive variables $(\rho, Wv^i, p)$ with the same method for GRHD described in \cite{2020CQGra..37n5015C,2021MNRAS.tmp.2378C} by removing the electromagnetic part.

\section{\label{sec:divb}Divergence-free handling for magnetic fields}
The time-component of equation~\eqref{eq:maxwell_eq2} implies that the divergence of the magnetic field is zero, namely, the solenoidal constraint~\eqref{eq:divb_0}:
\begin{equation}\tag{\ref{eq:divb_0}}
	\begin{aligned}
		&\nabla \cdot \vec{B} := \frac{1}{\sqrt{\gamma}} \partial_i \left( \sqrt{\gamma} B^i \right) = 0 \\
        	\Rightarrow & \hat{\nabla}_i q_{B^i} = \frac{1}{\sqrt{\hat{\gamma}}} \partial_i \left( \sqrt{\hat{\gamma}} q_{B^i} \right) = 0.
	\end{aligned}
\end{equation}
In practice, this condition is not held if we evolve the induction equation directly without any treatment due to the accumulating numerical error.
As a result, the non-vanishing monopoles are introduced and thus return non-physical results.
Various treatments are introduced to enforce this constraint in (general-relativistic) magnetohydrodynamics calculations.
Currently, three approaches are implemented, namely, (i) hyperbolic divergence cleaning (also known as generalised Lagrange multiplier (GLM)); (ii) (upwind) constrained transport schemes; and (iii) elliptic cleaning on cell-centred or face-centred meshes.

\subsection{\label{sec:glm}Generalised Lagrange multiplier} 
The generalised Lagrange multiplier (GLM) method, also known as hyperbolic divergence cleaning, is widely used in the community, e.g. \cite{DEDNER2002645,2009MNRAS.394.1727P,2013PhRvD..88d4020D,2021PhRvD.103d3022S}.
This method can also be applied to preserve charge conservation during the numerical evolution of electric fields $E^i$, which is also included in this work, will be discussed here.
In GLM method, the Maxwell and Faraday tensors are extended by scaler fields $\Psi$ and $\Phi$ to control the $\nabla \cdot \vec{B} = 0$ and charge conservation.
Maxwell equations \eqref{eq:maxwell_eq1} and \eqref{eq:maxwell_eq2} can then be extended as
\begin{align}
	\nabla_\nu \left( F^{\mu\nu} - \Psi g^{\mu\nu} \right) &= \mathcal{J}^\mu - \kappa_E n^\mu \Psi \label{eq:maxwell_eq1_psi}, \\
	\nabla_\nu \left( {}^*F^{\mu\nu} - \Phi g^{\mu\nu} \right) &= - \kappa_B n^\mu \Phi \label{eq:maxwell_eq2_phi}, 
\end{align}
where $\kappa_E$ and $\kappa_B$ are parameters.
Equations~\eqref{eq:maxwell_eq1_psi} and \eqref{eq:maxwell_eq2_phi} reduces to usual Maxwell equations if $\Psi = 0 = \Phi$.
The scalar fields $\Psi$ and $\Phi$ are being damped exponentially if $\kappa > 0$.
In other words, the standard Maxwell equations are reduced from the augmented Maxwell equations~\eqref{eq:maxwell_eq1_psi} and \eqref{eq:maxwell_eq2_phi} over a timescale $\sim 1/\kappa_E$ and $\sim 1/\kappa_B$.

To solve the augmented Maxwell equation, we need to include the evolution of the newly introduced scalar field $\Phi$.
This can be obtained by projecting equation~\eqref{eq:maxwell_eq1_psi} and \eqref{eq:maxwell_eq2_phi} to $ - n_\mu$:
\begin{align}
	&\begin{aligned}
	\frac{1}{\alpha \sqrt{\gamma}}&\left\{ 
	\partial_t \left( \sqrt{\gamma} \Psi \right) - \partial_i \left[ \sqrt{\gamma} \left( \alpha E^i -  \Psi \beta^i \right)\right] \right\} \\
	 & =  - \kappa_E \Psi + E^i \partial_i \alpha / \alpha + \Psi \gamma^{ij}K_{ij} + \rho_e,
	\end{aligned}\\
	&\begin{aligned}
	\frac{1}{\alpha \sqrt{\gamma}}&\left\{ 
	\partial_t \left( \sqrt{\gamma} \Phi \right) - \partial_i \left[ \sqrt{\gamma} \left( \alpha B^i -  \Phi \beta^i \right)\right] \right\} \\
	 & =  - \kappa_B \Phi + B^i \partial_i \alpha / \alpha + \Phi \gamma^{ij}K_{ij}.
	\end{aligned}
\end{align}
To be adopted in the reference metric form, we define:
\begin{align}
	q_\Psi :=& \psi^6 \sqrt{\bar{\gamma}/\hat{\gamma}} \Psi ,\\
	(f_\Psi)^i :=&  \psi^6 \sqrt{\bar{\gamma}/\hat{\gamma}} \left( \alpha {E}^i - \Psi \beta^i \right) 
	= \alpha q_{E^i} - q_\Psi \beta^i 	,\\
	s_\Psi :=& \psi^6 \sqrt{\bar{\gamma}/\hat{\gamma}} \left[ \alpha \rho_e + \alpha \Psi \left( \gamma^{ij} K_{ij} - \kappa_E \right) + E^i\partial_i \alpha \right],
\end{align}
and
\begin{align}
	q_\Phi :=& \psi^6 \sqrt{\bar{\gamma}/\hat{\gamma}} \Phi ,\\
	(f_\Phi)^i :=&  \psi^6 \sqrt{\bar{\gamma}/\hat{\gamma}} \left( \alpha {B}^i - \Phi \beta^i \right) 
	= \alpha q_{B^i} - q_\Phi \beta^i 	,\\
	s_\Phi :=& \psi^6 \sqrt{\bar{\gamma}/\hat{\gamma}} \left[ \alpha \Phi \left( \gamma^{ij} K_{ij} - \kappa_B \right) + B^i\partial_i \alpha \right] .
\end{align}
In addition, due to the existence of scalar fields $\Psi$ and $\Phi$, the evolution equation of the electric and magnetic fields $E^i$ and $B^i$ need to be modified as well.
The source terms of the evolution equations for electric and magnetic fields are modified as 
\begin{align}
	s_{E^i} :=& \psi^6 \sqrt{\bar{\gamma}/\hat{\gamma}} \left( \rho_e \beta^i - \alpha J^i  - \alpha \gamma^{ij} \partial_j \Psi \right), \\
	s_{B^i} :=& \psi^6 \sqrt{\bar{\gamma}/\hat{\gamma}} \left( - \alpha \gamma^{ij} \partial_j \Phi \right) .
\end{align}
Note that the modified Faraday equations are not hyperbolic since the existence of $ \partial_j \Psi$ and $ \partial_j \Phi$.

\subsection{\label{sec:ct}Constrained transport}
Constrained transport (CT) scheme, proposed by Evans and Hawley \cite{1988ApJ...332..659E}, is a robust divergence-control approach, in which the solenoidal constraint is kept in the machine precision level.
In this approach, the magnetic field $B^i$ are defined at the \emph{cell interface}.
By integrating the induction equation for each surface of the cell, together with Stokes' theorem, we have, for example
\begin{equation}\label{eq:B_surface}
	\begin{aligned}
		& \frac{\partial}{\partial t} \int_{\partial V \left(x^1_\texttt{i$\pm$1/2,j,k}\right) } \sqrt{\gamma}B^1_\texttt{i$\pm$1/2,j,k} dx^2 dx^3 \\ 
		& = \oint_{\partial A \left(x^1_\texttt{i$\pm$1/2,j,k}\right) } \hat{E}_k dx^k,
	\end{aligned}
\end{equation}
which can be written as
\begin{align}\label{eq:ct}
	&\frac{d}{dt} \Phi_\texttt{i$\pm$1/2,j,k} = \mathcal{E}_\texttt{i$\pm$1/2,j,k}.
\end{align}
Here, we have defined the magnetic fluxes
\begin{equation}
	\begin{aligned}
		\Phi_\texttt{i$\pm$1/2,j,k} & = \int_{\partial V \left(x^1_\texttt{i$\pm$1/2,j,k}\right) } \sqrt{\gamma}B^1_\texttt{i$\pm$1/2,j,k} dx^2 dx^3 \\
		& \approx \left( q_{B^1} \Delta A^1 \right) \Big|_\texttt{i$\pm$1/2,j,k}, 
	\end{aligned}
\end{equation}
while the electromotive forces (EMFs) $\mathcal{E}$ are, for example,
\begin{equation}
	\begin{aligned}
		& \mathcal{E}_\texttt{i$\pm$1/2,j,k} \approx\\
		& - \left[ \left({\hat{E}_3} \Delta x^3 \right)\Big|_\texttt{i$\pm$1/2,j+1/2,k} - \left({\hat{E}_3} \Delta x^3 \right)\Big|_\texttt{i$\pm$1/2,j-1/2,k} \right] \\
		& + \left[ \left({\hat{E}_2} \Delta x^2 \right)\Big|_\texttt{i$\pm$1/2,j,k+1/2} - \left({\hat{E}_2} \Delta x^2 \right)\Big|_\texttt{i$\pm$1/2,j,k-1/2} \right]
		,
	\end{aligned}
\end{equation}
where $\Delta A$ is the surface area while $\Delta x$ is the grid size of the cell (see \cite{2021MNRAS.tmp.2378C}).

As long as the ``electric field'' $\hat{E}^i$ at \emph{cell edges} are found, the magnetic field $B^i$ can be updated correspondingly by evaluating equation~\eqref{eq:ct}.
Note that, in this case, all the primitive variables are defined at \emph{cell centre} expect that the magnetic field are defined at \emph{cell interface}, some interpolation or reconstruction are needed to obtain $\hat{E}^i$ at \emph{cell edges}.
The electromagnetic fields arise in the advection terms in evolution equations, the reconstruction adopted here must also be upwind-type.
In \texttt{Gmunu}, the reconstruction based on HLL Riemann solver of $\hat{E}^i$ has been implemented, which will be discussed in the following.

\subsubsection{Upwind constrained transport based on HLL Riemann solver}
The first upwind constrained transport method was proposed in \cite{2004JCoPh.195...17L} in which limited reconstructions (such as WENO, MP5, etc.) is used.
Unlike arithmetic averaging method, upwind constrained transport method reduces to the correct one-dimensional limit when another two directions are symmetric.
The implementation of upwind constrained transport follows \cite{2004JCoPh.195...17L}.
The steps are the following:
\begin{enumerate}[(i)]
	\item Store the characteristic speeds $c^{-}_i$ and $c^{+}_i$ form the Riemann solver when calculating the fluxes at the interfaces, note that here $i=1,2,3$ represent the direction, 
	\item Reconstruct the left and right states of the electric field $E_3^{\rm{L}}$ and $E_3^{\rm{R}}$ from cell-centred to cell-interfaces.
	\item Reconstruct the left and right states of the electric field together with the lapse function $E_3^{\rm{LL}}$, $E_3^{\rm{LR}}$, $E_3^{\rm{RL}}$ and $E_3^{\rm{RR}}$ from cell-interfaces to cell-edges.
	\item Reconstruct the conformally rescaled magnetic fields to the cell edge. 
		This returns $q_{B^{1,2}}^{L,R}$.
	\item Calculate the electric field $\hat{E}_3$ at the \emph{cell edge} by
		\begin{equation}
			\begin{aligned}
				& \hat{E}_3^{\rm{LL}} = \alpha E_3^{\rm{LL}} + \sqrt{\hat{\gamma}} \left( \beta^1 q_{B^2}^{\rm{L}} - \beta^2 q_{B^1}^{\rm{L}} \right), \\
				& \hat{E}_3^{\rm{LR}} = \alpha E_3^{\rm{LR}} + \sqrt{\hat{\gamma}} \left( \beta^1 q_{B^2}^{\rm{L}} - \beta^2 q_{B^1}^{\rm{R}} \right), \\
				& \hat{E}_3^{\rm{RL}} = \alpha E_3^{\rm{RL}} + \sqrt{\hat{\gamma}} \left( \beta^1 q_{B^2}^{\rm{R}} - \beta^2 q_{B^1}^{\rm{L}} \right), \\
				& \hat{E}_3^{\rm{RR}} = \alpha E_3^{\rm{RR}} + \sqrt{\hat{\gamma}} \left( \beta^1 q_{B^2}^{\rm{R}} - \beta^2 q_{B^1}^{\rm{R}} \right),
			\end{aligned}
		\end{equation}
		where $\sqrt{\hat{\gamma}}$ can be determined analytically while the metric variables such as $\alpha$ and $\beta^i$ are obtained via interpolation instead of reconstruction.
	\item Take the maximum characteristic speeds in each direction from the four states
		\begin{equation}
			\begin{aligned}
				& c_{1}^{-} = \max \left( c_{1,\rm{L}}^{-}, c_{1,\rm{R}}^{-} \right), \\
				& c_{1}^{+} = \max \left( c_{1,\rm{L}}^{+}, c_{1,\rm{R}}^{+} \right), \\
				& c_{2}^{-} = \max \left( c_{2,\rm{L}}^{-}, c_{2,\rm{R}}^{-} \right), \\
				& c_{2}^{+} = \max \left( c_{2,\rm{L}}^{+}, c_{2,\rm{R}}^{+} \right),
			\end{aligned}
		\end{equation}
		where $c_{i,\rm{L/R}}^{-} \left(c_{i,\rm{L/R}}^{+} \right)$ are the non-negative characteristic speeds in the $-i (+i)$ direction.
	\item Estimate the electric field $\hat{E}_3$ at the cell edge by the formula proposed in \cite{2004JCoPh.195...17L}:
		\begin{equation}
			\begin{aligned}
				& \hat{E}_3 = \\ 
					\Bigg[
					& \frac{ c_{1}^{+}c_{2}^{+}\hat{E}_3^{\rm{LL}} 
						+ c_{1}^{+}c_{2}^{-}\hat{E}_3^{\rm{LR}} 
						+ c_{1}^{-}c_{2}^{+}\hat{E}_3^{\rm{RL}} 
						+ c_{1}^{-}c_{2}^{-}\hat{E}_3^{\rm{RR}} 
						}{\left(c_{1}^{+}+c_{1}^{-}\right)\left(c_{2}^{+}+c_{2}^{-}\right)} \\
					& + \frac{c_{1}^{+}c_{1}^{-}}{c_{1}^{+}+c_{1}^{-}} \left( q_{B^2}^{\rm{R}} - q_{B^2}^{\rm{L}} \right)
					- \frac{c_{2}^{+}c_{2}^{-}}{c_{2}^{+}+c_{2}^{-}} \left( q_{B^1}^{\rm{R}} - q_{B^1}^{\rm{L}} \right)
				\Bigg].
			\end{aligned}
		\end{equation}
\end{enumerate}
Note that as discussed in \cite{2007A&A...473...11D}, this HLL based upwind constrained transport scheme in ideal-magnetohydrodynamics can be simplified.
The details of the implementation in this case can be found in appendix~\ref{sec:uct_ideal}

\subsection{\label{sec:elliptic_cleaning}Elliptic cleaning}
The so-called \emph{elliptic} divergence cleaning, works by solving Poisson's equation and enforcing a divergence-free magnetic field:
\begin{align}
        & \hat{\nabla}^2 \Phi = \hat{\nabla}_i q_{B^i}^{\text{old}} , \label{eq:poisson}\\
	& q_{B^i}^{\text{new}} = q_{B^i}^{\text{old}} - \left( \hat{\nabla}\Phi \right)^i \label{eq:update_bfield}.
\end{align}
Whenever the conserved magnetic field $q_{B^i}$ is updated at each timestep, we first solve Poisson's equation \eqref{eq:poisson} through the multigrid solver, then we update the magnetic field with the solution $\Phi$ as shown in equation~\eqref{eq:update_bfield}.

In previous work \cite{2021MNRAS.tmp.2378C}, we present the first GRMHD code which adopts elliptic divergence cleaning as the divergence-free treatment during the evolution in the literature.
This scheme was developed only for the magnetic field is defined at cell centres.
Although elliptic was shown working properly in various tests, this scheme is technically acausal in relativistic setting.
Using this scheme to control divergence-free constrain only could be problematic in some extreme cases where the magnetic monopole gains rapidly at each timestep.
Indeed, this scheme can be applied in stagger meshes naturally, which can be used together with constrained transport schemes to further suppress the amplitude of magnetic monopoles in a causal manner.

\section{Magnetic fields initialisation}
To satisfy the solenoidal constraint~\eqref{eq:divb_0} initially, we compute the magnetic fields $B^i$ from the vector potential $A_i$.
In particular, the magnetic fields can be obtained from the vector potential by
\begin{equation}
	\begin{aligned}
		& B^i = \epsilon^{ijk} \partial_j A_k \\
		\Rightarrow & q_{B^i} = \frac{1}{\sqrt{\hat{\gamma}}} \eta^{ijk} \partial_j A_k.
	\end{aligned}
\end{equation}
Second order finite differencing are used to compute the curl of the vector potential (defined at cell centres) when the magnetic fields are defined at cell-centres.
On the other hand, when stagger-grid magnetic fields are being used, we compute the magnetic fields in a way that similar with constraint transport method (see section~\ref{sec:ct}) by defining the vector potential $A_i$ at the cell-edges.

Additionally, on top of obtaining the magnetic fields from the vector potential $A_i$, we further suppress the violation of the solenoidal constraint by applying the elliptic cleaning (see section~\ref{sec:elliptic_cleaning}) at the beginning of the simulations.

\section{Numerical tests}\label{sec:numerical_tests}
In the remainder of this paper, we present a selection of representative test problems with our code.
The tests {range} from special relativistic magnetohydrodynamics to general relativistic magnetohydrodynamics, from {one to multiple dimensions}.
In practice, we do not observed any strong violation of charge conservation among the simulations we have.
Therefore, the charge conservation is not handled by default.
Unless otherwise specified, all simulations in flat spacetime reported in this paper were performed with Harten, Lax and van Leer (HLL) Riemann solver \cite{harten1983upstream}, 2-nd order Montonized central (MC) limiter \cite{1974JCoPh..14..361V} with IMEX-SSP2(2,2,2) time integrator \cite{pareschi2005implicit}.

For the simulations of neutron stars, we used the same refinement setting as in our preivous work \cite{2021MNRAS.tmp.2378C}.
In particular, we defined a relativistic gravitational potential $\Phi := 1 - \alpha$.
For any $\Phi$ larger than the maximum potential $\Phi_{\text{max}}$ (which is set as 0.2 in this work), the block is set to be finest.
While for the second-finest level, the same check is performed with a new maximum potential which is half of the previous one, so on and so forth.
The grid is updated every 500 timesteps.
In these neutron star tests, IMEXCB3a time stepper \cite{2015JCoPh.286..172C} and PPM reconstruction are used.

\subsection{Magnetic monopole test for hyperbolic cleaning}
One way to assess the hyperbolic cleaning scheme is to evolve a system which magnetic monopoles are included artificially.
To add the Gaussian monopoles, by following \cite{2014CQGra..31a5005M}, consider 
\begin{align}
	B^x(r) = 
	\begin{cases}
		\exp \left[ - r^2 / R_\text{G}^2 \right] - \exp\left( - 1 \right)	&	\text{if } r < R_\text{G} ,\\
		0	&	\text{otherwise},
	\end{cases}
\end{align}
where $R_\text{G}$ is the radius of the Gaussian monopoles, which is set as $0.2$ is this work.
In this test, we assume the matter is uniformly distributed in the entire computational domain, with the resistivity $\eta = 10^{-16}$.
In particular, we set $\rho = 1$, $\epsilon = 0.1$, $v^i=0$ in this test.
We consider an ideal-gas equation of state $ p = (\Gamma-1)\rho \epsilon$ with $\Gamma = 5/3$.
The computational domain covers $[-2,2]$ for both $x$, $y$ and $z$ directions with the resolution $64 \times 64 \times 64$.
To see how the hyperbolic cleaning parameter $\kappa$ affects the evolution, we choose $\kappa=1$, $5$ and $10$.

\begin{figure}
	\centering
	\includegraphics[width=\columnwidth, angle=0]{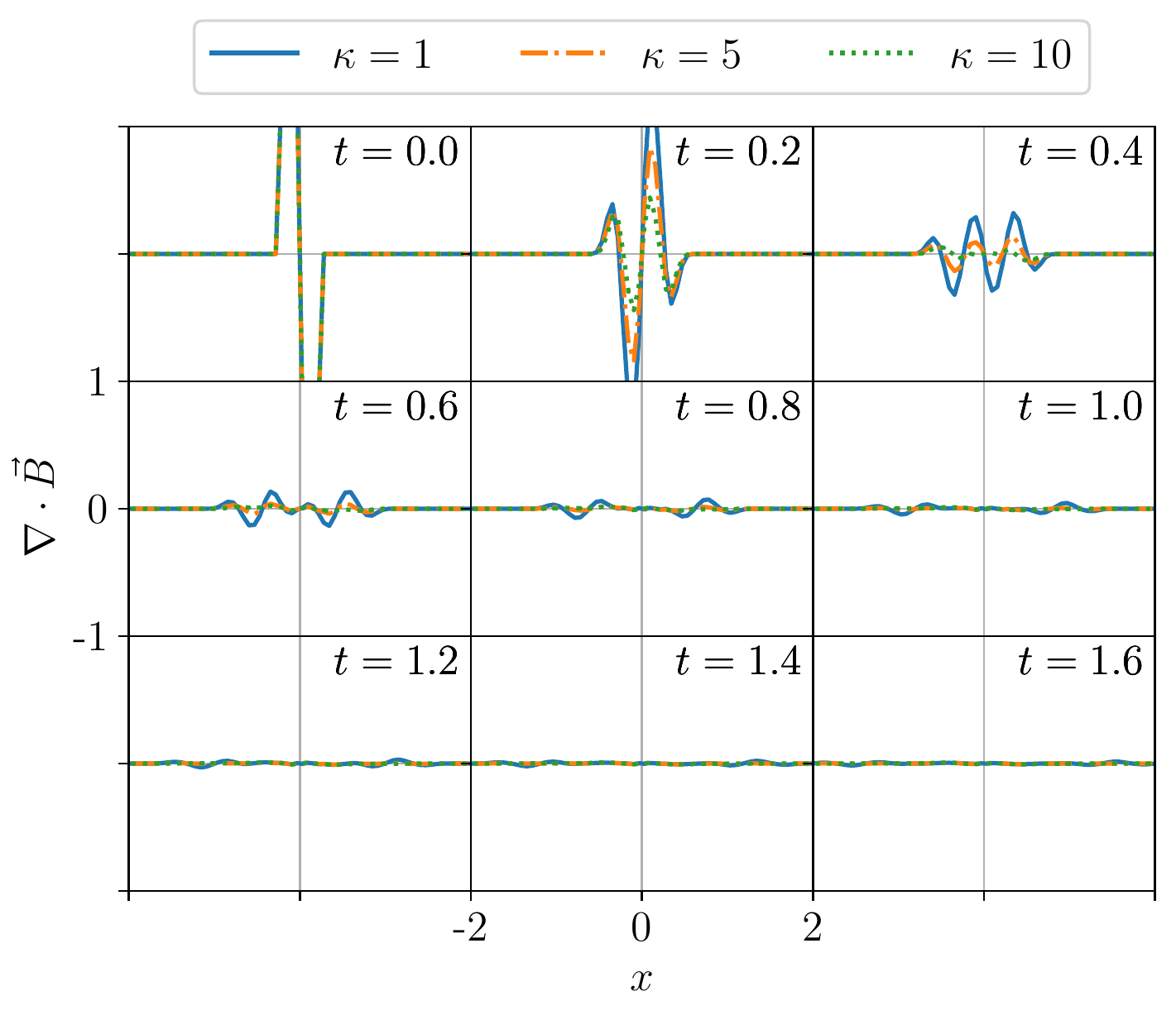}
	\caption{
		Comparison of the magnetic monopole $\nabla \cdot \vec{B}$ with different $\kappa$ at different time steps.
		As expected, the damping rate increase as $\kappa$ increases.
		}
	\label{fig:EMHD_monopole_test}	
\end{figure}

Figure~\ref{fig:EMHD_monopole_test} compares of the magnetic monopole $\nabla \cdot \vec{B}$ with different $\kappa$ at different time steps.
As expected, the damping rate increase as $\kappa$ increases.

\subsection{Relativistic Shock Tubes}
Standard tests for assessing the capacity to capture shocks are shock tube tests.
In particular, a test for ideal relativistic MHD is presented in \cite{2001ApJS..132...83B}.
Here we follow the same setup, but with non-vanishing resistivity.
In particular, we perform the simulation with {Cartesian} coordinates on {a} flat spacetime.
The initial condition is given as
\begin{align}
	\left( \rho, p, B^x, B^y \right) = 
	\begin{cases}
		\left( 1,1,0.5,1 \right)	&	\text{if } x < 0 ,\\
		\left( 0.125,0.1,0.5,-1 \right)	&	\text{if } x > 0.
	\end{cases}
\end{align}
We consider an ideal-gas equation of state $ p = (\Gamma-1)\rho \epsilon$ with $\Gamma = 2$.
In this test, we vary the resistivity $\eta$ from $10^{-6}$ to $10^{3}$.

\begin{figure}
	\centering
	\includegraphics[width=1.0\columnwidth, angle=0]{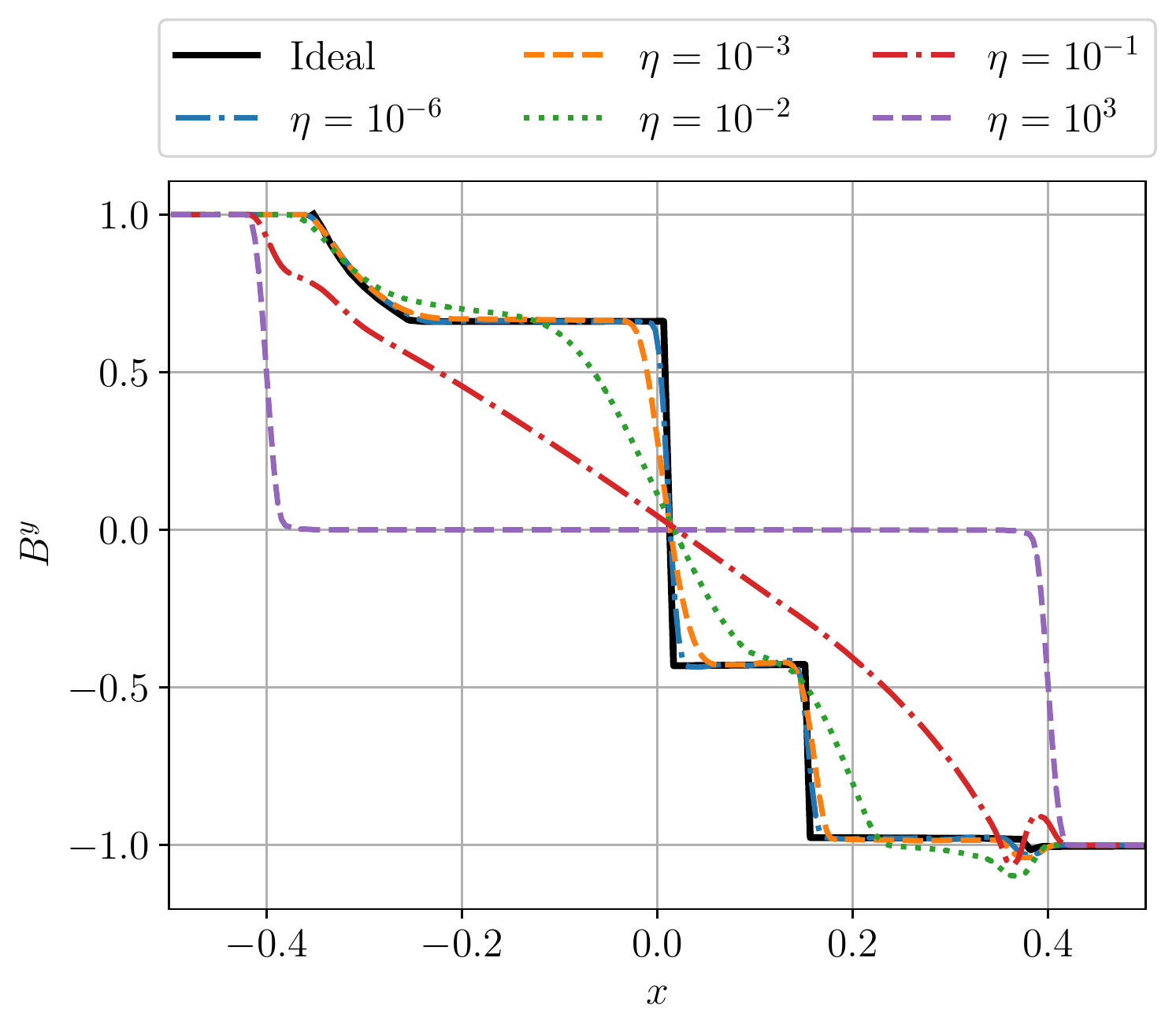}
	\caption{Figure shows the $y$-component of the magnetic field $B^y$ for the shock tube test at $t = 0.4$ with the resistivity in range $\eta \in \left[ 10^{-6}, 10^{3}\right]$.
		With the low resistivity e.g., $\eta \leq 10^{-3}$, the result is close to the ideal MHD case ($\eta \rightarrow 0$, black solid line).
		}
	\label{fig:EMHD_shocktube_1d}	
\end{figure}

Figure~\ref{fig:EMHD_shocktube_1d} compares the numerical results obtained by \texttt{Gmunu} with the reference solutions (black solid lines) \citet{2001ApJS..132...83B} of the $y$-component of the magnetic field $B^y$ at $t=0.4$.

\subsection{Self-similar current sheet}
The self-similar current sheet test describes the self-similar evolution of a thin current sheet, was first considered in \cite{2007MNRAS.382..995K}.
The analytic expression of the $y$ component of the magnetic field $B^y$ and the $z$ component of the electric field $E^z$ is given by
\begin{align}
	B^y(x,t) &= \textrm{erf}\left( \frac{ x }{2\sqrt{\eta t}} \right), \\
	E^z(x,t) &= \sqrt{\frac{\eta}{\pi t}}\exp\left( -\frac{x}{4{\eta t}} \right). 
\end{align}
This test is preformed by setting $t=1$ as initial condition, and we set the resistivity $\eta = 0.01$.
The rest-mass density and pressure are uniformly distributed with $\rho=1$ and $p = 5000$ respectively. 
The computational domain covers $[-1.5,1.5]$ for $x$.

\begin{figure}
	\centering
	\includegraphics[width=\columnwidth, angle=0]{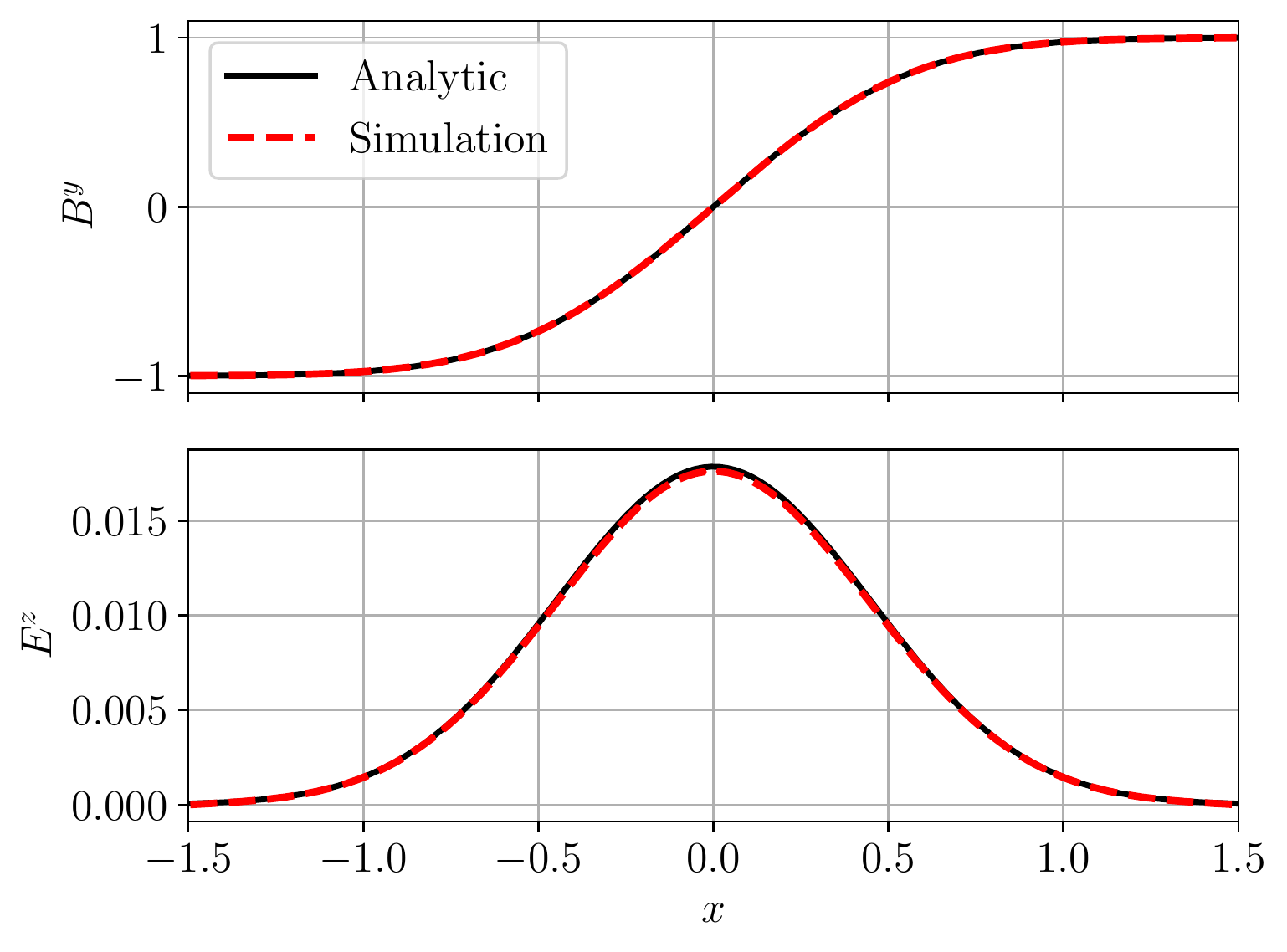}
	\caption{ 
		The figure shows the simulated magnetic field $B^y$ and electric field $E^z$ (red dashed lines) at $t=10$ for the self-similar current sheet test problem with resolution $N=256$.
		The solid black lines are the analytic solutions.
		The numerical results obtained by \texttt{Gmunu} agree with the analytic solutions.
		}
	\label{fig:EMHD_self_similar_current_sheet}	
\end{figure}

Figure~\ref{fig:EMHD_self_similar_current_sheet} compares the analytic solutions and the simulated results of the $y$ component of the magnetic field $B^y$ and the $z$ component of the electric field $E^z$ at $t=10$ with resolution $N=256$.
The numerical results obtained by \texttt{Gmunu} agree with the analytic solutions.

\begin{figure*}
	\begin{minipage}{.48\textwidth}
		\centering
		\begin{tabular}{c|c|c}
			\hline
			\hline
			$N$  & $\delta_N$ & $R_N$ \\ \hline
			32   & 9.237E-3   & --    \\ 
			64   & 1.646E-4   & 2.48  \\ 
			128  & 2.638E-4   & 2.64  \\ 
			256  & 4.609E-5   & 2.52  \\ 
			512  & 8.995E-6   & 2.36  \\ 
			1024 & 1.914E-6   & 2.23  \\ 
			2048 & 4.197E-7   & 2.19  \\
			4096 & 8.056E-8   & 2.38  \\
			\hline
			\hline
		\end{tabular}
	\end{minipage}
	\begin{minipage}{0.48\textwidth}
		\includegraphics[width=\columnwidth, angle=0]{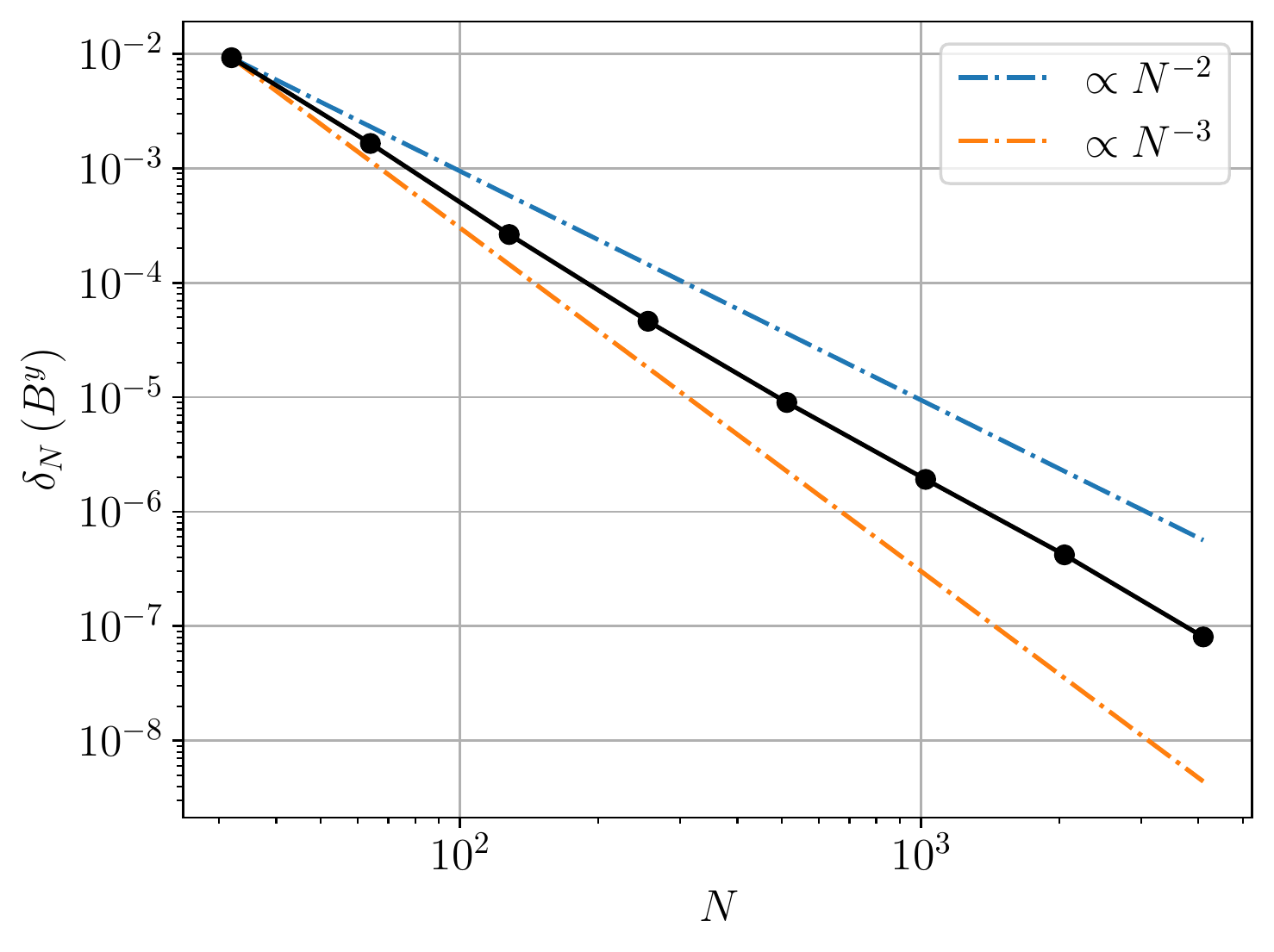}
	\end{minipage}
	\caption{
		Numerical errors versus resolution for the self-similar current sheet test problem at $t=10$.
		In this test, 2-nd order accurate strong-stability preserving IMEX-SSP2(2,2,2) time integrator \cite{pareschi2005implicit}, Harten, Lax and van Leer (HLL) Riemann solver \cite{harten1983upstream} with 2-nd order Montonized central (MC) limiter \cite{1974JCoPh..14..361V} are used. 
		\emph{Left panel}: The table of numerical errors of versus resolution for this problem.
		\emph{Right panel}: The numerical errors of $B^y$ versus resolution (black line) for the self-similar current sheet problem.
		Second-order ideal scaling is given by the dashed blue line while third-order ideal scaling is given by the dashed orange line.
		The convergence rate is between second-order and third-order in this test.
		}
	\label{fig:EMHD_self_similar_current_sheet_convergence}	
\end{figure*}

While the analytical solution is available in this smooth test, additional care is needed if we want to use it to assess the convergence rate.
As discussed in \cite{2013MNRAS.428...71B}, the exact solution valid only in the case of infinite pressure $p \rightarrow \infty$.
Hence, instead of comparing with the exact solution, we follow the approach presented in \cite{2013MNRAS.428...71B} by using a high-resolution run as our reference solution.
The relative error of the $y$ component of the magnetic field $\delta_N\left(B^y\right)$ is defined as
\begin{equation}
	\begin{aligned}
		& \delta_N \left( B^{y}(t = T_{\text{final}})\right) := \\
		& \frac{1}{N} \sum\limits_{i=1}^{N} \frac{\lvert B^{y}(t = T_{\text{final}}) - B^{y}_{\text{ref}}(t = T_{\text{final}}) \rvert}{\max\left(B^{y}(t = T_{\text{final}})\right)},
	\end{aligned}
\end{equation}
and the convergence rate $R_N$ can be obtained by
\begin{equation} \label{eq:convergence_rate}
	R_N = \log_2 \left( \frac{\delta_{N/2}}{\delta_N} \right).
\end{equation}
In our study, the reference solution of $B^y$ is provided by a simulation with the resolution $ N = 8192$.
Figure~\ref{fig:EMHD_self_similar_current_sheet_convergence} shows the numerical errors of $B^y$ versus resolution $N$ (blue line) for this problem.
Second-order ideal scaling is given by the dashed black line while third-order ideal scaling is given by the dashed orange line.
The convergence rate is between second-order and third-order in this test.

\subsection{Cylindrical blast wave}
The cylindrical blast wave is a well-known challenging multi-dimensional SRMHD test problem, which describes an expanding blast wave in a plasma with an initially uniform magnetic field.
The initial condition of this test problem is determined with radial parameters $r_\text{in}$ and $r_\text{out}$.
The density (and also the pressure, in the same form) profile is given by:
\begin{align}
	\rho(r) = 
	\begin{cases}
		\rho_\text{in}	&	\text{if } r \leq r_\text{in} ,\\
		\exp \left[ \frac{ \left(r_\text{out} - r\right) \ln \rho_\text{in} +  \left(r - r_\text{in}\right) \ln \rho_\text{out} }{r_\text{out} - r_\text{in}} \right]	&	\text{if }  r_\text{in} \leq r \leq r_\text{out} ,\\
		\rho_\text{out}	&	\text{if } r \geq r_\text{out} ,
	\end{cases}
\end{align}
where the parameters are: 
\begin{align}
	& r_\text{in} = 0.8, && r_\text{out} = 1.0; \\
	& \rho_\text{in} = 10^{-2}, && \rho_\text{out} = 10^{-4} ;  \\
	& p_\text{in} = 1.0, && p_\text{out} = 5 \times 10^{-4};  \\
	& B^i = (0.1, 0, 0), && v^i = (0,0,0) .
\end{align}
Here we consider the ideal-gas equation of state with $\Gamma = 4/3$.
In this test, upwind constrained transport together with elliptical cleaning is used.
To recover to ideal MHD limit, the resistivity is set as $\eta = 10^{-16}$ in this test.
The computational domain covers $[-6,6]$ for both $x$ and $y$ directions with the resolution {$256 \times 256$}.

\begin{figure}
	\centering
	\includegraphics[width=1.0\columnwidth, angle=0]{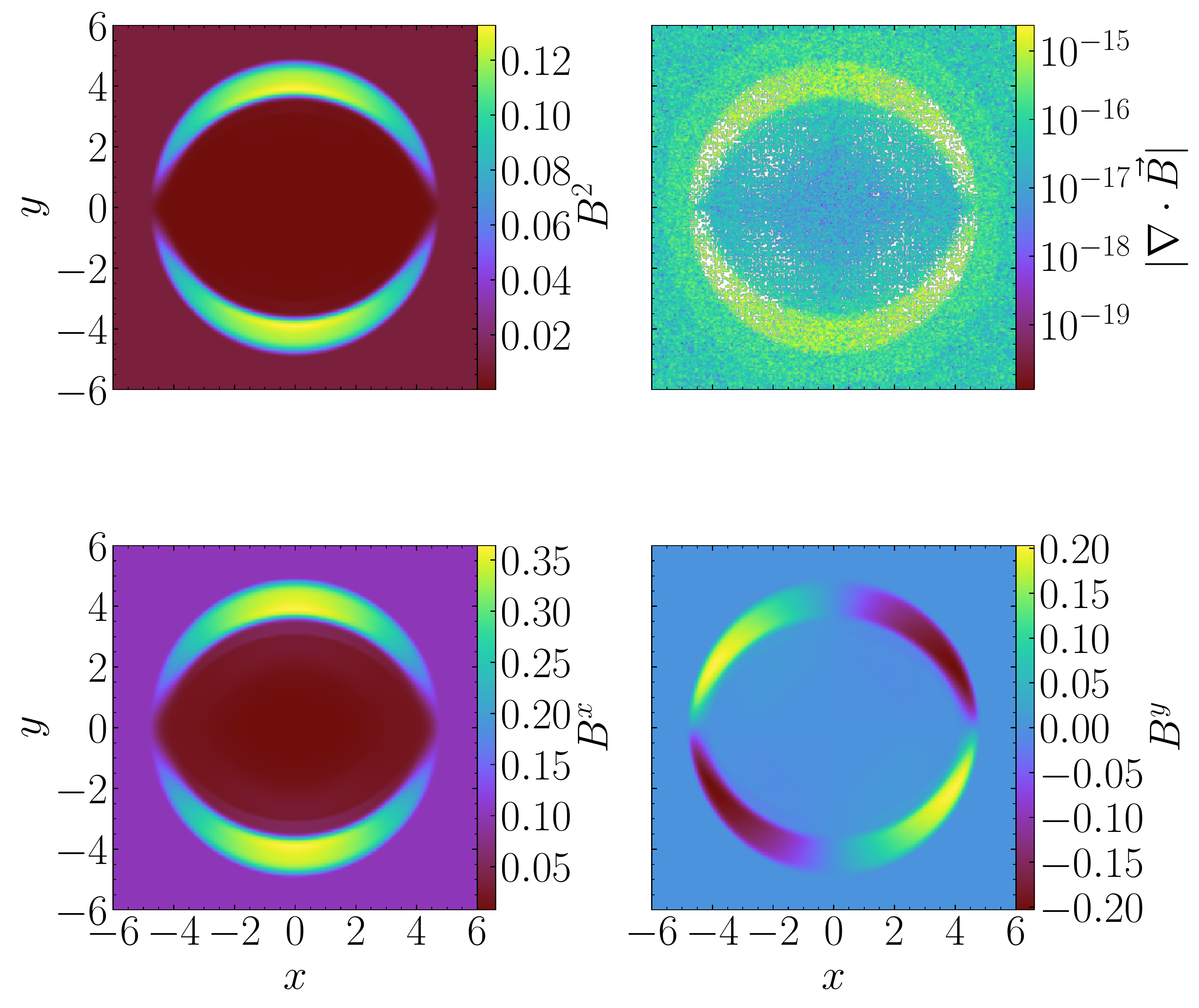}
	\caption{The two-dimensional profile for the cylindrical blast wave of the magnetic field strength $B^i B_i$ (\emph{upper left}), $\nabla \cdot \vec{B}$(\emph{upper right}), $B^x$(\emph{lower left}), $B^y$(\emph{lower right}) at $t = 4.0$.
		}
	\label{fig:EMHD_cylindrical_blast_wave_2d}	
\end{figure}

\begin{figure}
	\centering
	\includegraphics[width=1.0\columnwidth, angle=0]{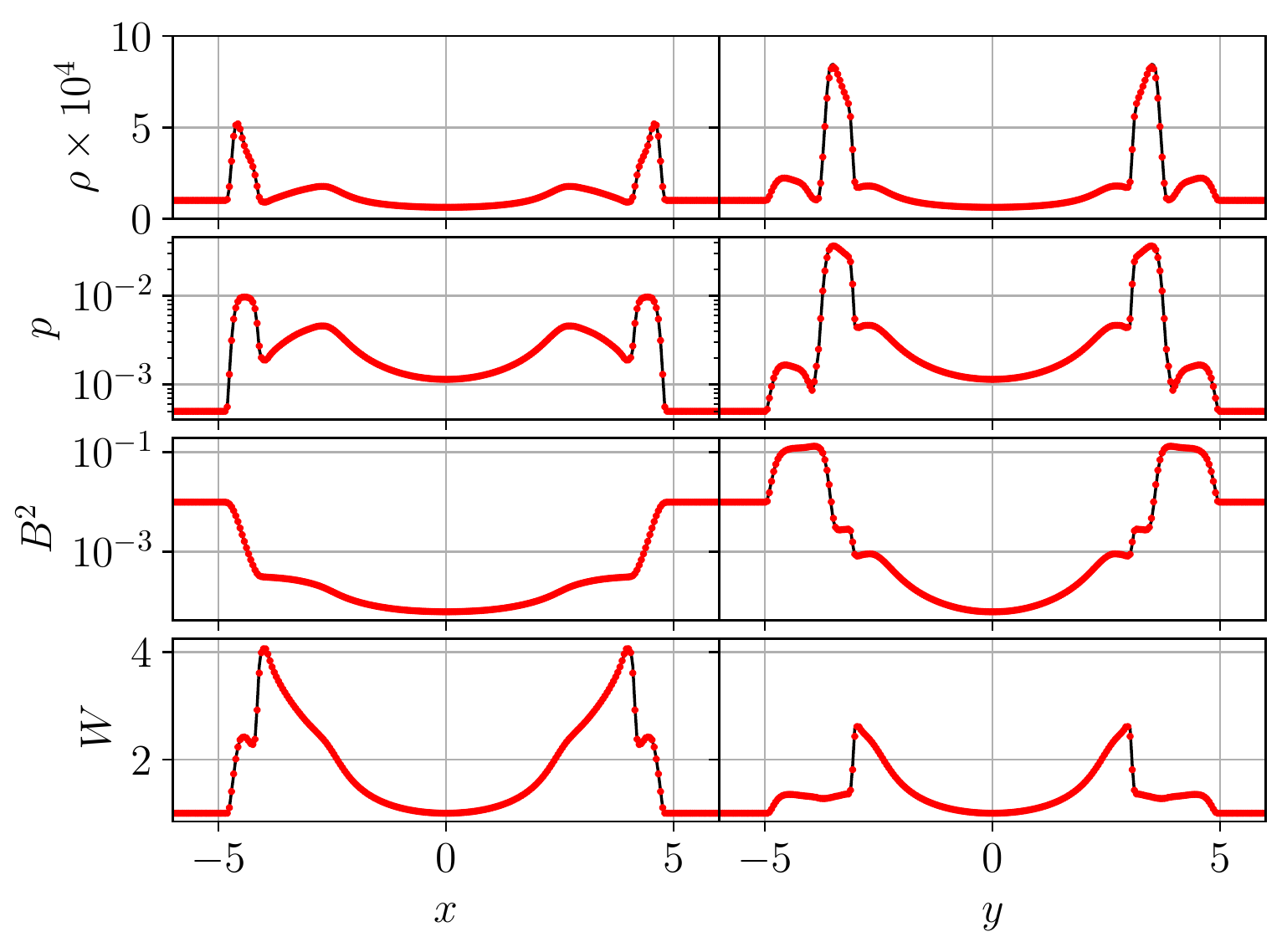}
	\caption{One-dimensional slices along the $x$-axis (\emph{left {column}}) and $y$-axis (\emph{right {column}}) for the density $\rho$ (\emph{{top row}}), {pressure} $p$ (\emph{{second row}}), $B^2$ (\emph{{third row}}) and Lorentz factor $W$ (\emph{{fourth row}}) for the MHD cylindrical blast wave test at $t = 4.0$.
	The red dots show the numerical results by solving resistive MHD equations while the black solid lines show the results by solving ideal MHD equations.
	There is no visible differences between ideal and resistive MHD cases.
		}
	\label{fig:EMHD_cylindrical_blast_wave_2d_slices}	
\end{figure}

Figure~\ref{fig:EMHD_cylindrical_blast_wave_2d} shows the two-dimensional profile of the magnetic field strength $B^i B_i $, $B^x$, $B^y$ and the divergence of the magnetic field $\nabla \cdot \vec{B}$ at $t = 4.0$. 
To compare the results with ideal MHD cases, we also plot one-dimensional slices along the $x-$ and $y-$ axes for the rest mass density $\rho$, pressure $p$, $B^2$ and the Lorentz factor $W$ at $t=4$, in figure~\ref{fig:EMHD_cylindrical_blast_wave_2d_slices}.
The numerical results obtained by resistive MHD agree with the ideal MHD approximation.

\subsection{Telegraph equation}
The propagation of the electromagnetic waves in a material with finite conductivity $\sigma_{\rm{c}}$ can be described by Maxwell's equations in the fluid rest frame,
\begin{equation} \label{eq:telegraph_eq_1}
	\begin{aligned}
		&\frac{\partial \vec{B}}{\partial t} + \nabla \times \vec{E} = 0 \\
		&\frac{\partial \vec{E}}{\partial t} - \nabla \times \vec{B} = - \sigma_{\rm{c}} \vec{E},
	\end{aligned}
\end{equation}
the solution of which also satisfies the telegraph equation
\begin{equation} \label{eq:telegraph_eq_2}
	\begin{aligned}
		&\frac{\partial^2 \vec{B}}{\partial t^2} + \sigma_{\rm{c}}\frac{\partial \vec{B}}{\partial t} = \nabla^2 \vec{B} \\
		&\frac{\partial^2 \vec{E}}{\partial t^2} + \sigma_{\rm{c}}\frac{\partial \vec{E}}{\partial t} = \nabla^2 \vec{E} - \nabla \rho_e. 
	\end{aligned}
\end{equation}
The dispersion relation of the plane wave solutions with wavenumber $k$ and frequency $\omega$ of equation~\eqref{eq:telegraph_eq_1} and the telegraph equation~\eqref{eq:telegraph_eq_2} is given by
\begin{equation}
	\omega = -i \frac{\sigma_{\rm{c}}}{2} \pm \mu, \qquad \rm{where } \qquad \mu = \sqrt{k^2 - \frac{\sigma_{\rm{c}}^2}{4}}.
\end{equation}
The exact solution of equation~\eqref{eq:telegraph_eq_1} can be expressed as
\begin{equation} \label{eq:telegraph_sol}
	\begin{aligned}
		B^z_{\rm{exact}} = & B_1 \exp\left(-\frac{\sigma_{\rm{c}}}{2} t\right) \cos\left( \vec{k} \cdot \vec{x} - \mu t \right)\\
		E^y_{\rm{exact}} = & B_1 \exp\left(-\frac{\sigma_{\rm{c}}}{2} t\right) \times \\
		& \left[ \frac{\mu}{k}\cos\left( \vec{k} \cdot \vec{x} - \mu t \right) + \frac{\sigma_{\rm{c}}}{2k}\sin\left( \vec{k} \cdot \vec{x} - \mu t \right) \right],
	\end{aligned}
\end{equation}
where $B_1$ is the initial perturbation amplitude.

To make this as a two-dimensional test, we follow the setting suggested in \cite{2019MNRAS.486.4252M}.
In particular, the computational domain is set as $L_x \times L_y = 1 \times 0.5$ with the resolution $N_x \times N_y = N_x \times N_x/2$ with periodic boundary condition.
The solution~\eqref{eq:telegraph_sol} is rotated along $z$-axis with an angle $\alpha$, 
\begin{equation}
	\left\{ \vec{E},\vec{B} \right\} = 
	\begin{bmatrix}
		\cos \alpha & -\sin\alpha & 0 \\
		\sin \alpha & \cos\alpha & 0 \\
		0 & 0 & 1 \\
	\end{bmatrix}
	\left\{ \vec{E}_{\rm{exact}},\vec{B}_{\rm{exact}} \right\}, 
\end{equation}
where $\tan \alpha = L_x/L_y = 2$, conductivity is $\sigma_{\rm{c}} = 1$ and the initial perturbation amplitude $B_1 = 1$.
The wavevector is set as $\vec{k} = k_x \left(1, \tan\alpha, 0\right)$, where $k_x = 2\pi / L_x$.
The rest-mass density and pressure are uniformly distributed with $\rho=10^{12}$ and $p = 0$ respectively, with zero fluid velocities.

\begin{figure*}
	\begin{minipage}{.48\textwidth}
		\centering
		\begin{tabular}{c|c|c}
			\hline
			\hline
			$N_x$  & $\lvert\lvert B^z(T) - B^z_{\rm{exact}}(T) \rvert\rvert_1$ & $R_N$ \\ \hline
			16   & 1.928E-1   & --    \\ 
			32   & 3.891E-2   & 2.31  \\ 
			64   & 1.162E-2   & 1.74  \\ 
			128  & 2.922E-3   & 1.99  \\ 
			256  & 5.969E-4   & 2.29  \\ 
			512  & 1.474E-4   & 2.02  \\ 
			1024 & 3.669E-5   & 2.01  \\ 
			2048 & 9.084E-6   & 2.01  \\
			4096 & 2.221E-6   & 2.03  \\
			\hline
			\hline
		\end{tabular}
	\end{minipage}
	\begin{minipage}{0.48\textwidth}
		\includegraphics[width=\columnwidth, angle=0]{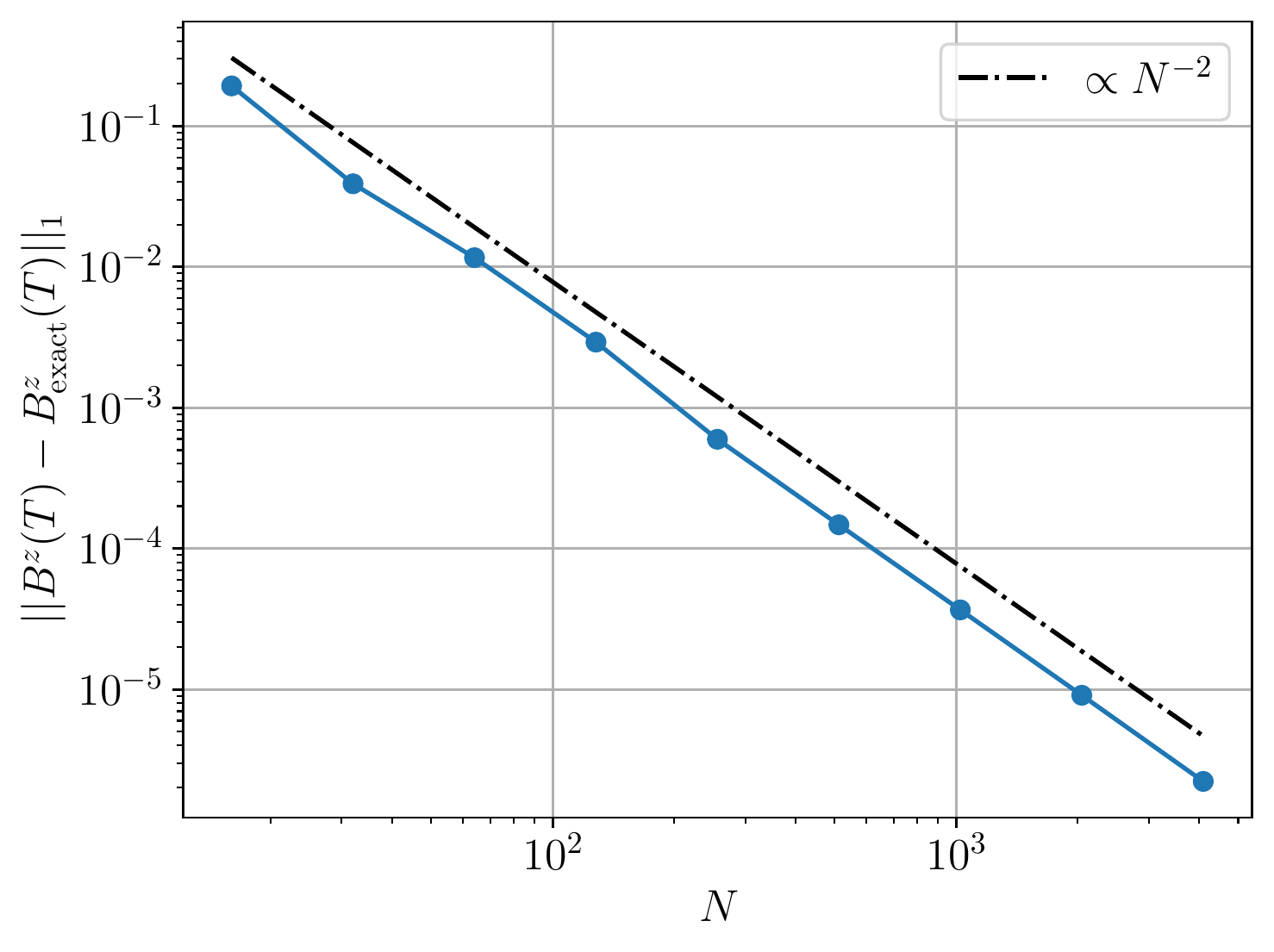}
	\end{minipage}
	\caption{
		Numerical errors versus resolution for the two-dimensional relativistic-resistive-magnetohydrodynamics telegraph equation test.
		In this test, 2-nd order accurate strong-stability preserving IMEX-SSP2(2,2,2) time integrator \cite{pareschi2005implicit}, Harten, Lax and van Leer (HLL) Riemann solver \cite{harten1983upstream} with 2-nd order Montonized central (MC) limiter \cite{1974JCoPh..14..361V} are used. 
		\emph{Left panel}: Table of the $L_1$-norm of the difference of the difference between the initial and final values of the $z$-component of the magnetic field $\lvert\lvert B^z(T) - B^z_{\rm{exact}}(T) \rvert\rvert_1$ and convergence rates of this problem at $t=T$ at different resolution $N$.
		\emph{Right panel}: The $L_1$-norm-versus-resolution plot. 
		Second-order ideal scaling is given by the dashed black line.
		As expected, the second-order convergence is achieved with this setting.
		}
	\label{fig:EMHD_telegraph_equation_convergence_rate}	
\end{figure*}

To quantify the convergence rate at $t=T=2\pi / \mu \approx 0.44749706611091833$, we compute the $L_1$-norm of the difference of the difference between the numerical and exact values of the $z$-component of the magnetic field $\lvert\lvert B^z(T) - B^z_{\rm{exact}}(T) \rvert\rvert_1$ and the convergence rate $R_N$ (see equation~\ref{eq:convergence_rate}).
The left panel of figure~\ref{fig:EMHD_telegraph_equation_convergence_rate} shows the $L_1$-norm of the difference of the difference between the initial and final values of the $z$-component of the magnetic field $\lvert\lvert B^z(T) - B^z_{\rm{exact}}(T) \rvert\rvert_1$ and convergence rates of this problem at $t=T$ at different resolution $N$.
The convergence rate can be virtually present with a numerical-errors-versus-resolution plot, as shown in the right panel of figure~\ref{fig:EMHD_telegraph_equation_convergence_rate}.
As expected, the second-order convergence is achieved with this setting.

\subsection{Stationary charged vortex}
The first exact two-dimensional test for resistive magneto-hydrodynamics was introduced recently \cite{2019MNRAS.486.4252M}, which describes a rotating flow with uniform density in a vertical magnetic field and a radial electric field.
This solution can be expressed in cylindrical coordinates $(R,z,\varphi)$ by
\begin{align}
	E_R = & \frac{q_0}{2} \frac{R}{R^2+1}, \\
	B_z = & \frac{\sqrt{\left(R^2 + 1 \right)^2 - q_0^2 / 4}}{R^2+1}, \\
	v_\varphi = & - \frac{q_0}{2} \frac{R}{\sqrt{\left(R^2 + 1 \right)^2 - q_0^2 / 4}}, \\
	p = & -\rho \frac{\Gamma-1}{\Gamma} \nonumber \\
		& + \left( p_0 + \rho \frac{\Gamma-1}{\Gamma}\right) 
	\left(\frac{4R^2 + 4 - q_0^2 }{ \left( R^2 + 1 \right)\left( 4 - q_0^2 \right)}\right)^{\frac{\Gamma}{2\left(\Gamma-1\right)}},
\end{align}
where we choose the charge density $q_0 = 0.7$, pressure $p_0 = 0.1$ and the density $\rho = 1$ in the whole domain.
We consider an ideal-gas equation of state $ p = (\Gamma-1)\rho \epsilon$ with the adiabatic index $\Gamma = 4/3$ and the resistivity is $\eta = 10^{-3}$ in this case.
Note that the charge density is
\begin{equation}
	\rho_e = \frac{q_0}{\left( R^2 + 1 \right)^2},
\end{equation}
which can be computed from the gradient of the electric field.
The computational domain covers $[-10,10]$ for both $x$ and $y$ directions with the resolution $N \times N$.
All values at the outer boundaries are kept fixed to maintain this stationary configuration.

\begin{figure}
	\centering
	\includegraphics[width=\columnwidth, angle=0]{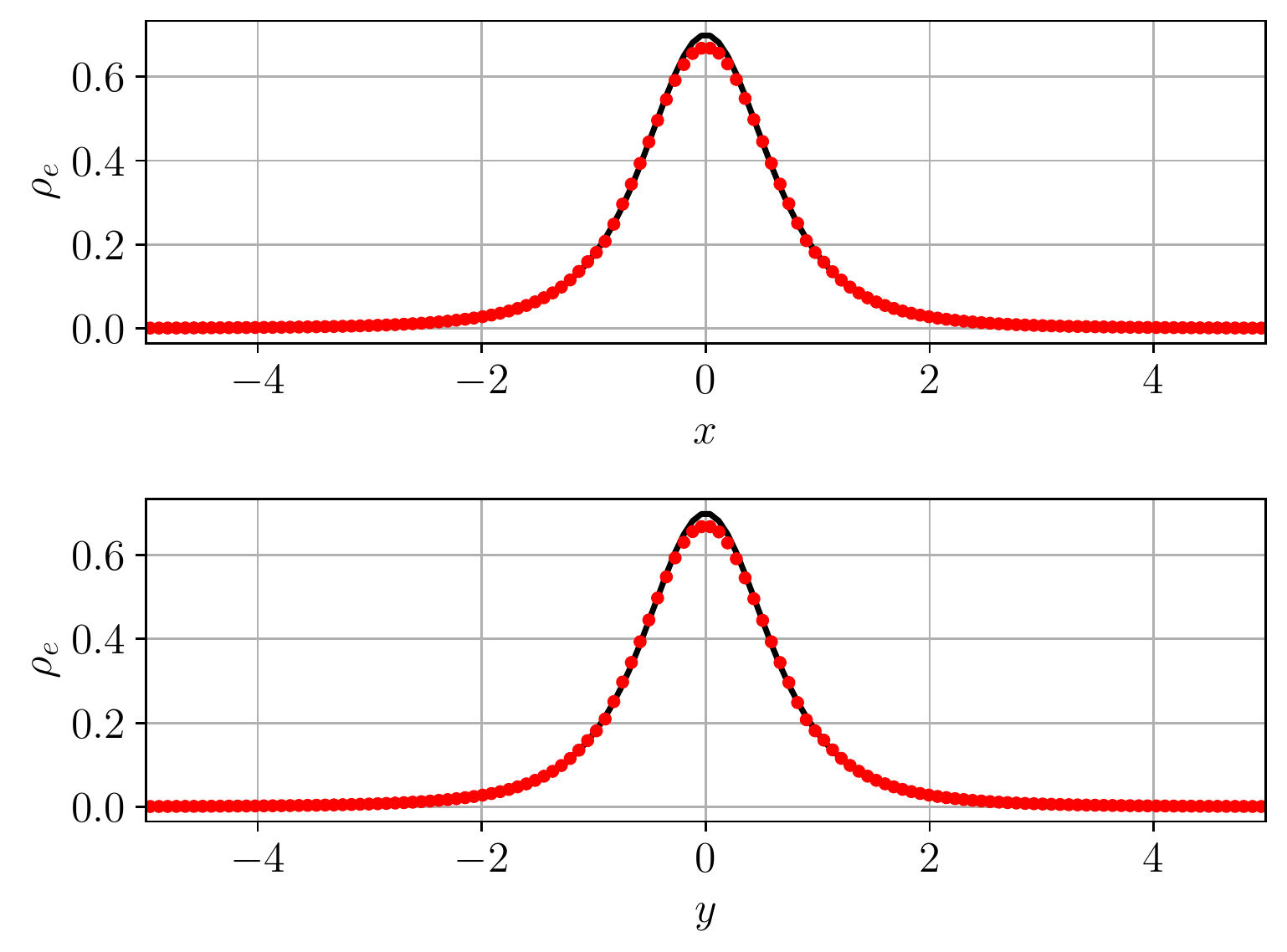}
	\caption{
		One-dimensional slices along the $x$-axis (\emph{upper panel}) and $y$-axis (\emph{lower panel}) for the charge density $\rho_e$ for the stationary charged vortex test at $t = 5$.
		The red dots show the numerical results obtained by \texttt{Gmunu}, which agree with the initial condition (black solid lines).
		}
	\label{fig:EMHD_charged_vortex}	
\end{figure}

Figure~\ref{fig:EMHD_charged_vortex} shows the one-dimensional slices along the $x$- and $y$-axes for the charge density $\rho_e$ at $t=5$.
The result shows that \texttt{Gmunu} is able to maintain the charge density $\rho_e$ well up to $t=5$.
\begin{figure}
	\centering
	\includegraphics[width=\columnwidth, angle=0]{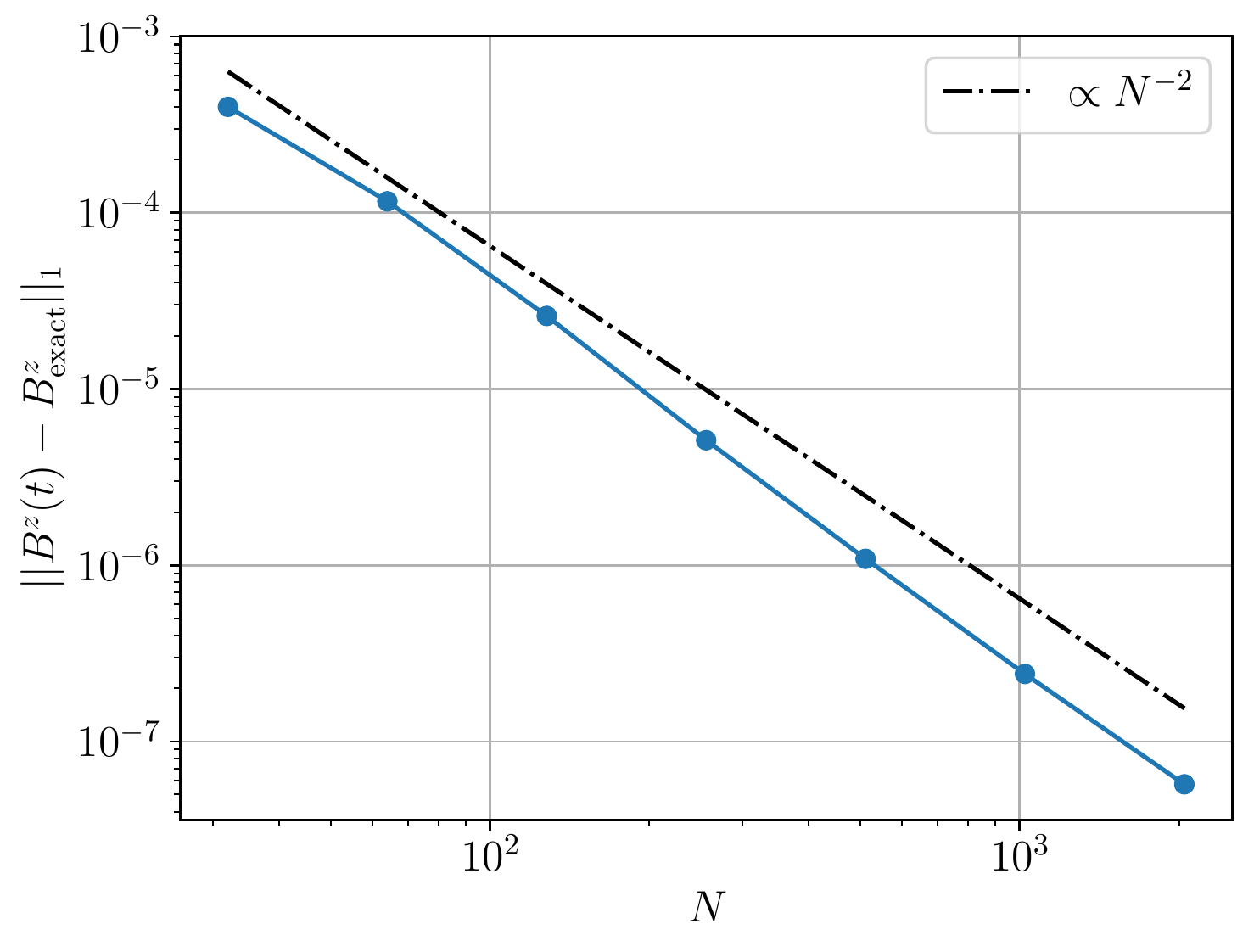}
	\caption{ 
		The $L_1$-norm of the difference of the difference between the initial and final values of the $z$-component of the magnetic field $\lvert\lvert B^z(t=10) - B^z(t=0) \rvert\rvert_1$ of the charged vortex problem at different resolution $N$.
		Second-order ideal scaling is given by the dashed black line.
		The convergence rate is between second-order and third-order in this test.
		}
	\label{fig:EMHD_charged_vortex_convergence}	
\end{figure}
\begin{table}
	\centering
	\begin{tabular}{c|c|c}
		\hline
		\hline
		$N$  & $L_1$-norm & $R_N$ \\ \hline
		32   & 3.990E-4   & --    \\ 
		64   & 1.161E-4   & 1.78  \\ 
		128  & 2.597E-5   & 2.16  \\ 
		256  & 5.122E-6   & 2.34  \\ 
		512  & 1.087E-6   & 2.24  \\ 
		1024 & 2.417E-7   & 2.17  \\ 
		2048 & 5.716E-8   & 2.08  \\
		\hline
		\hline
	\end{tabular}
	\caption{\label{tab:EMHD_charged_vortex_convergence_rate}
		Table of the $L_1$-norm of the difference of the difference between the initial and final values of the $z$-component of the magnetic field $\lvert\lvert B^z(t=10) - B^z(t=0) \rvert\rvert_1$ and convergence rates of the charged vortex problem at different resolution $N$.
		Numerical errors of $B^z$ and convergence rates for the self-similar current sheet test problem at $t=10$.
		In this test, HLL Riemann solver for the flux, MC reconstruction and IMEX-SSP2(2,2,2) for the time integration \cite{pareschi2005implicit} are used.
		As expected, the second-order convergence is achieved with this setting.
		}
\end{table}

\subsection{One-dimensional steady dynamo test}
Here we consider the case when the dynamo effect is taken into account by following \cite{2013MNRAS.428...71B}.
This test describes the growth of the magnetic field in a stationary medium.
The medium is filled with uniform density $\rho = 10^{-14}$ with zero velocities.
In this test, we freeze the evolution of the matter, only the evolution equations of electromagnetic fields are solved.
The initial condition of the electromagnetic fields are the following:
\begin{equation}
	\begin{aligned}
		B^x = 0, \qquad 
		B^y = 0.1 \sin \left( k x \right), \qquad
		B^z = - 0.1 \cos \left( k x \right), \qquad
	\end{aligned}
\end{equation}
and $E^i = 0$, where $k$ is the wavenumber, which is set as $1$ or $5$ in our tests.
The resistivity $\eta = 0.01$ and the dynamo parameter $\xi = 0.5$ are kept fixed.
The computational domain covers $[-\pi,\pi]$ for $x$ resolution $N_x=256$ with periodic boundary condition.
Analytic solution of this test is available (see appendix A in \cite{2013MNRAS.428...71B}), where the exponential growth rate of the magnetic field is 0.385 and 0.59 when $k=1$ and 5, respectively.

\begin{figure}
	\centering
	\includegraphics[width=\columnwidth, angle=0]{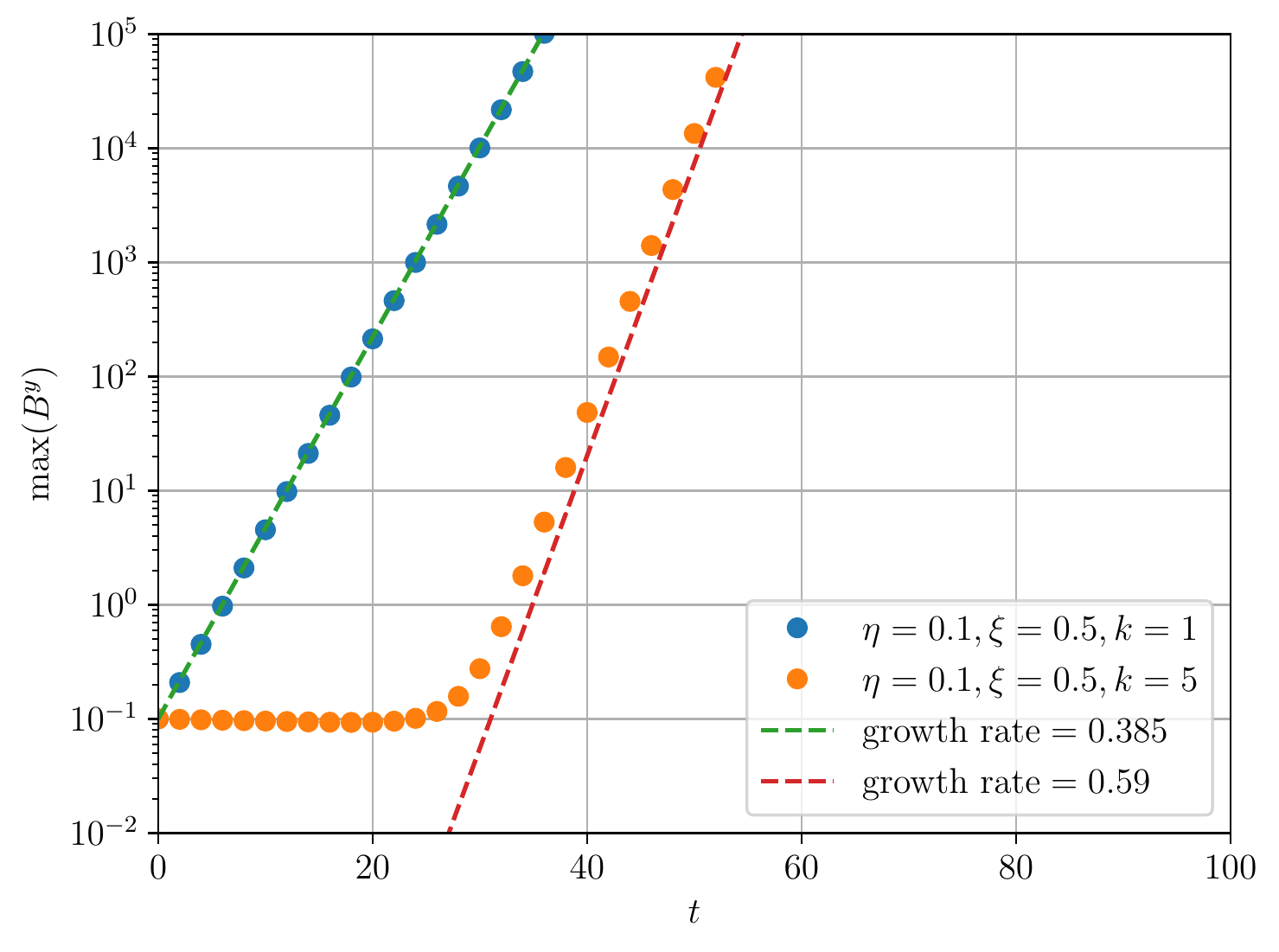}
	\caption{
		The evolution of the $y$-component of the magnetic field $B^y$ in one-dimensional steady dynamo test.
		The blue dots represent the exponential growth of $B^y$ when $k=1$, which is consistent with the analytical solution with the growth rate $0.385$ (green dashed line).
		In addition, the effects of resistivity and dynamo are cancelled by each other when $k=5$, the magnetic field is expected to remain unchanged in this case.
		However, due to numerical errors, the $y$-component of the magnetic field $B^y$ with $k=5$ (orange dots) is remained at the same order of magnitude up to $t \sim 20$ and eventually grows with the fastest growing mode with the rate $0.59$ (red dashed line).
		}
	\label{fig:EMHD_steady_dynamo}	
\end{figure}

Figure~\ref{fig:EMHD_steady_dynamo} shows the evolution of the $y$-component of the magnetic field $B^y$.
The blue dots represent the exponential growth of $B^y$ when $k=1$, which is consistent with the analytical solution with the growth rate $0.385$ (green dashed line).
In addition, the effects of resistivity and dynamo are cancelled by each other when $k=5$, the magnetic field is expected to remain unchanged in this case.
However, due to numerical errors, the $y$-component of the magnetic field $B^y$ with $k=5$ (orange dots) is remained at the same order of magnitude up to $t \sim 20$ and eventually grows with the fastest growing mode with the rate $0.59$ (red dashed line).

\subsection{Loop advection}
The advection of a weakly magnetised loop is a well known test to examine divergence-control technique in a magnetohydrodynamics code.
To focus on how the divergence-control affects the advection of this weakly magnetised loop, we simulate the problem in ideal MHD limit (by setting the resistivity is set as $\eta = 10^{-6}$).
The computational domain is set to be \emph{periodic} at all boundaries and covers the region $-1 \leq x \leq 1$ and $-0.5 \leq y \leq 0.5$ with the base grid points $n_x \times n_y = 32 \times 16$ and allowing 5 AMR {levels} (i.e., an effective resolution of $512 \times 256 $).
Note that in this test, the refinement is determined based on the strength of the magnetic field.
In particular, the grid is refined if the square of the magnetic field $B^iB_i$ is larger than $10^{-10}$ while it is coarsened otherwise.
This test is performed on a uniform background with $\rho = 1$, $p = 1$, $v^x=0.2$ and $v^y = 0.1$.
The initial condition of the magnetic field $B^i$ is given by the vector potential $A_i$
\begin{align}
	A_z = 
	\begin{cases}
		A_0 \left( R - r \right)	&	\text{if } r < R ,\\
		0	&	\text{if } r > R ,
	\end{cases}
\end{align}
where $R=3$ is the radius of the advecting magnetic loop, $r := \sqrt{x^2 + y^2}$ and $A_0$ is chosen to be $10^{-3}$.
We consider an ideal-gas equation of state $ p = (\Gamma-1)\rho \epsilon$ with $\Gamma = 4/3$.
With this setting, the period of this problem is $T = 10$, the simulations are preformed till $t_{\text{final}} = 100 = 10 T$.

\begin{figure*}
  \centering
  \begin{tabular}{@{}lcc@{}}
	  {} & $B^2$  & $|\nabla \cdot \vec{B}|$ \\
	  \rotatebox{90}{Hyperbolic cleaning} &
    \includegraphics[height=.15\textheight]{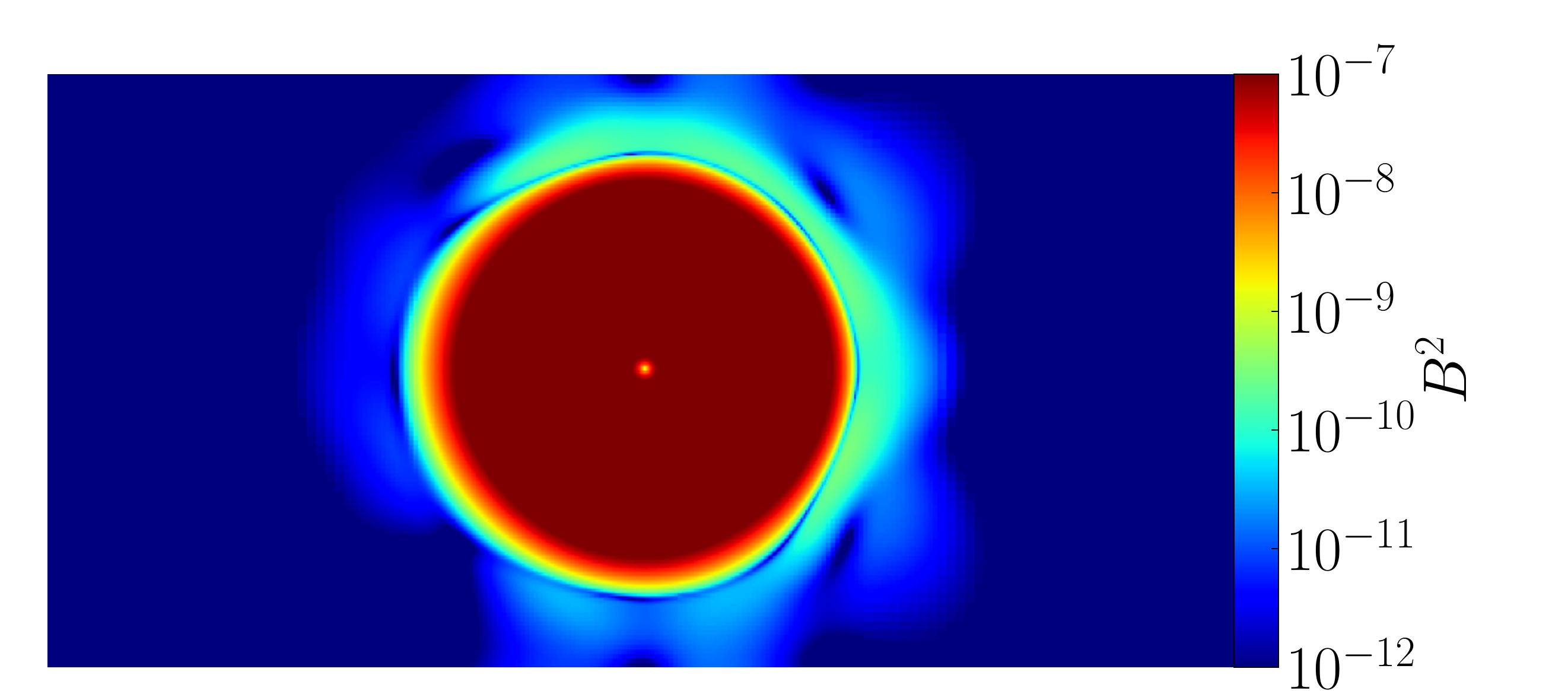} &
    \includegraphics[height=.15\textheight]{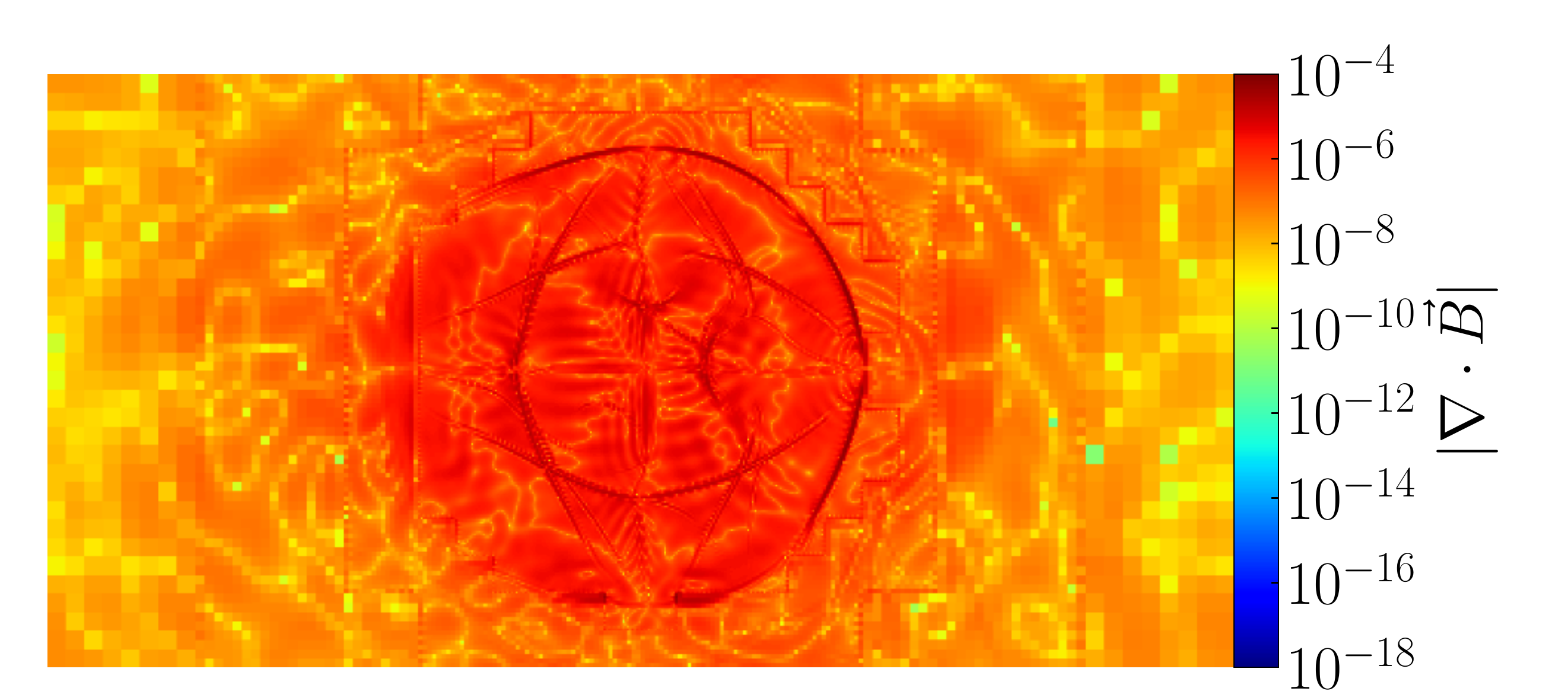} \\
	  \rotatebox{90}{Elliptic cleaning} &
    \includegraphics[height=.15\textheight]{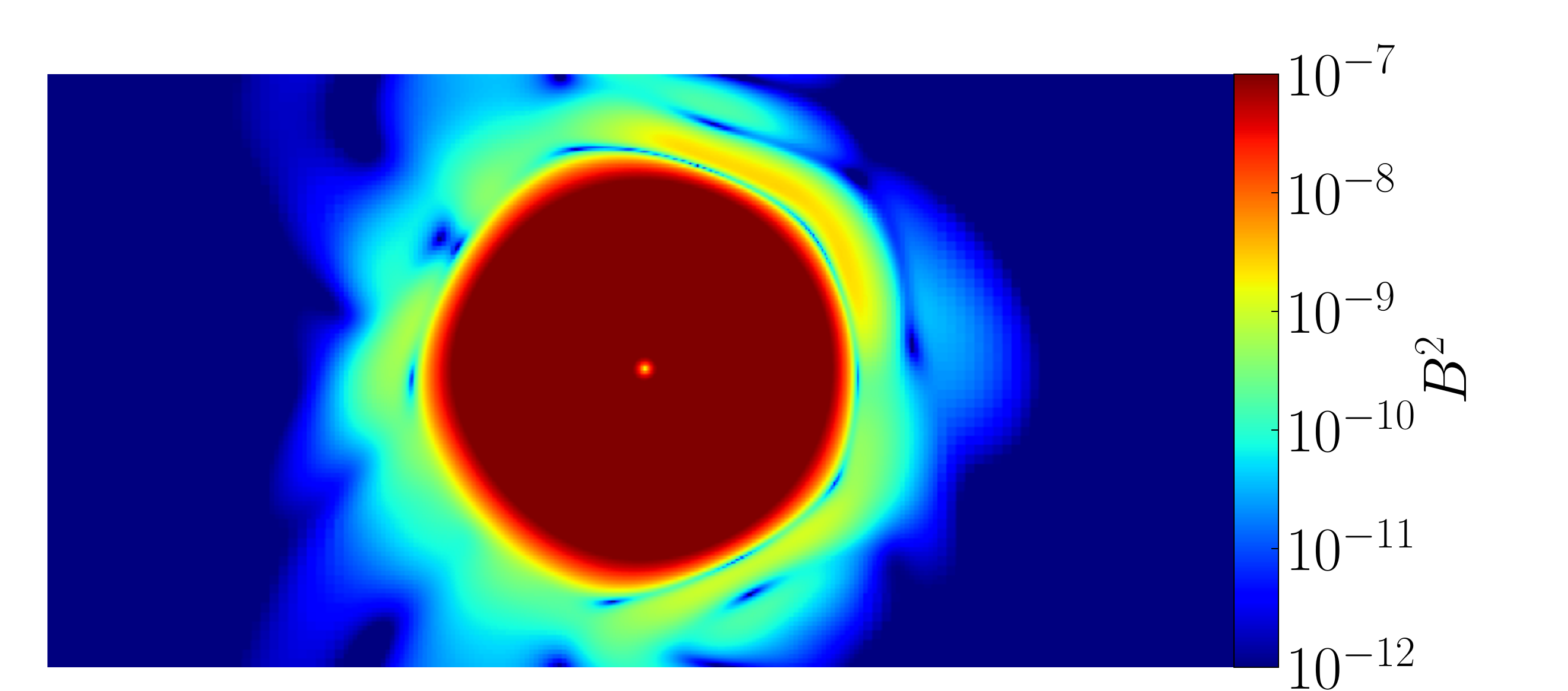} &
    \includegraphics[height=.15\textheight]{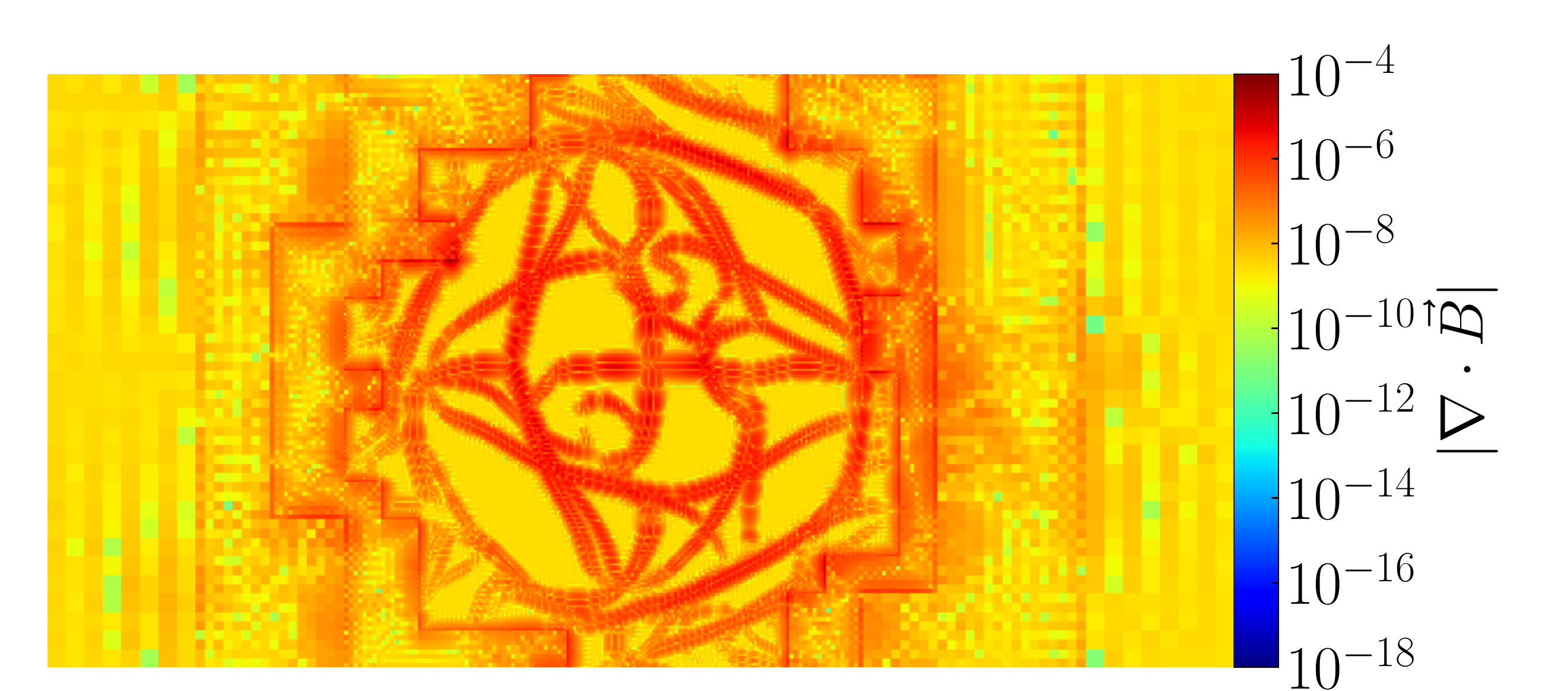} \\
	  \rotatebox{90}{UCT: HLL} &
    \includegraphics[height=.15\textheight]{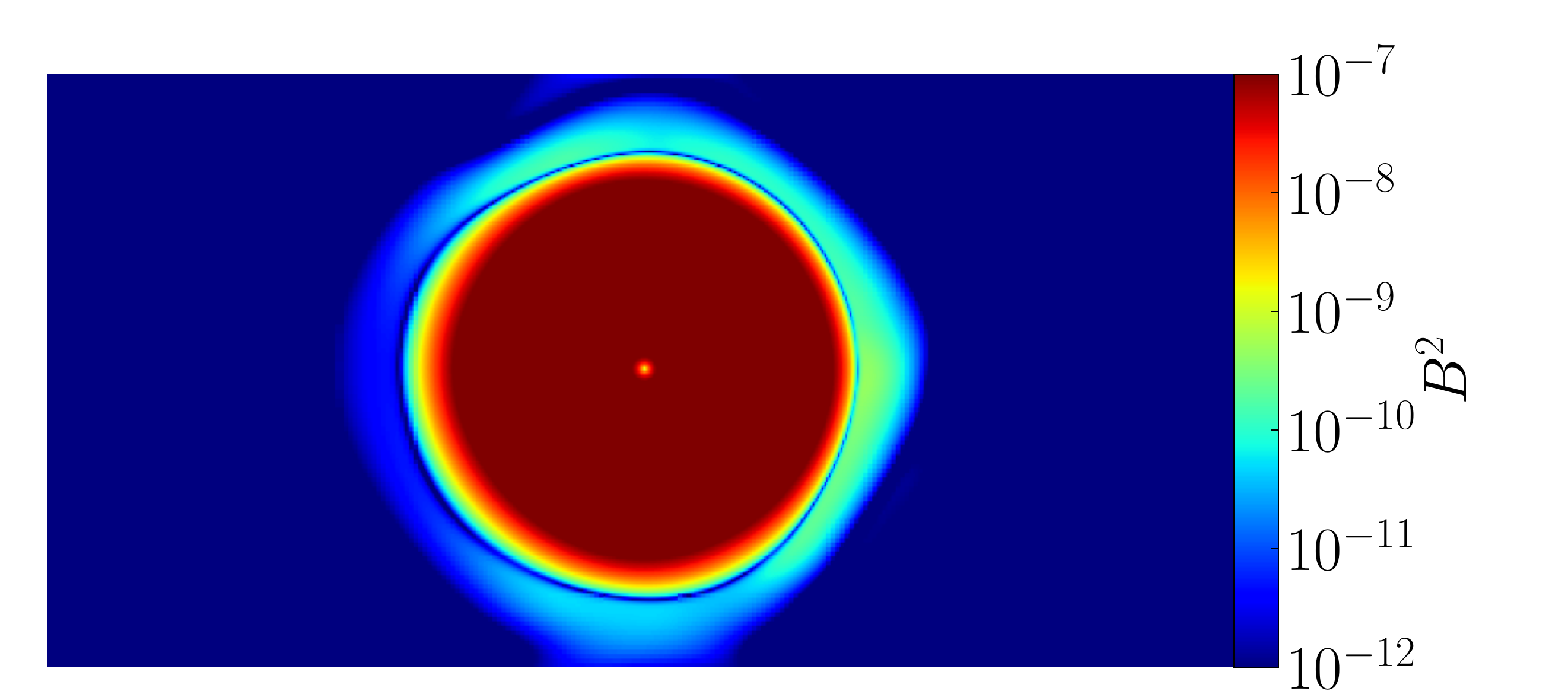} &
    \includegraphics[height=.15\textheight]{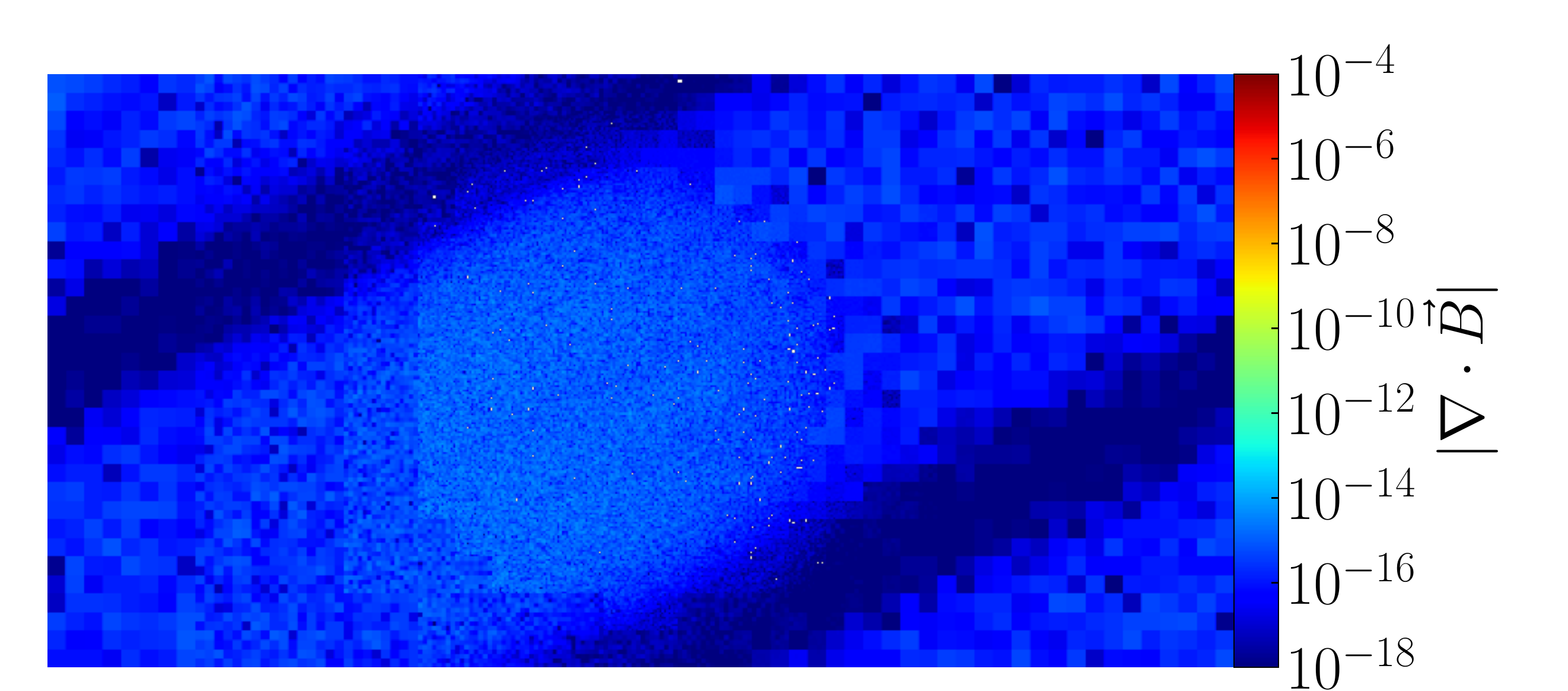} \\
	  \rotatebox{90}{UCT: HLL + MG} &
    \includegraphics[height=.15\textheight]{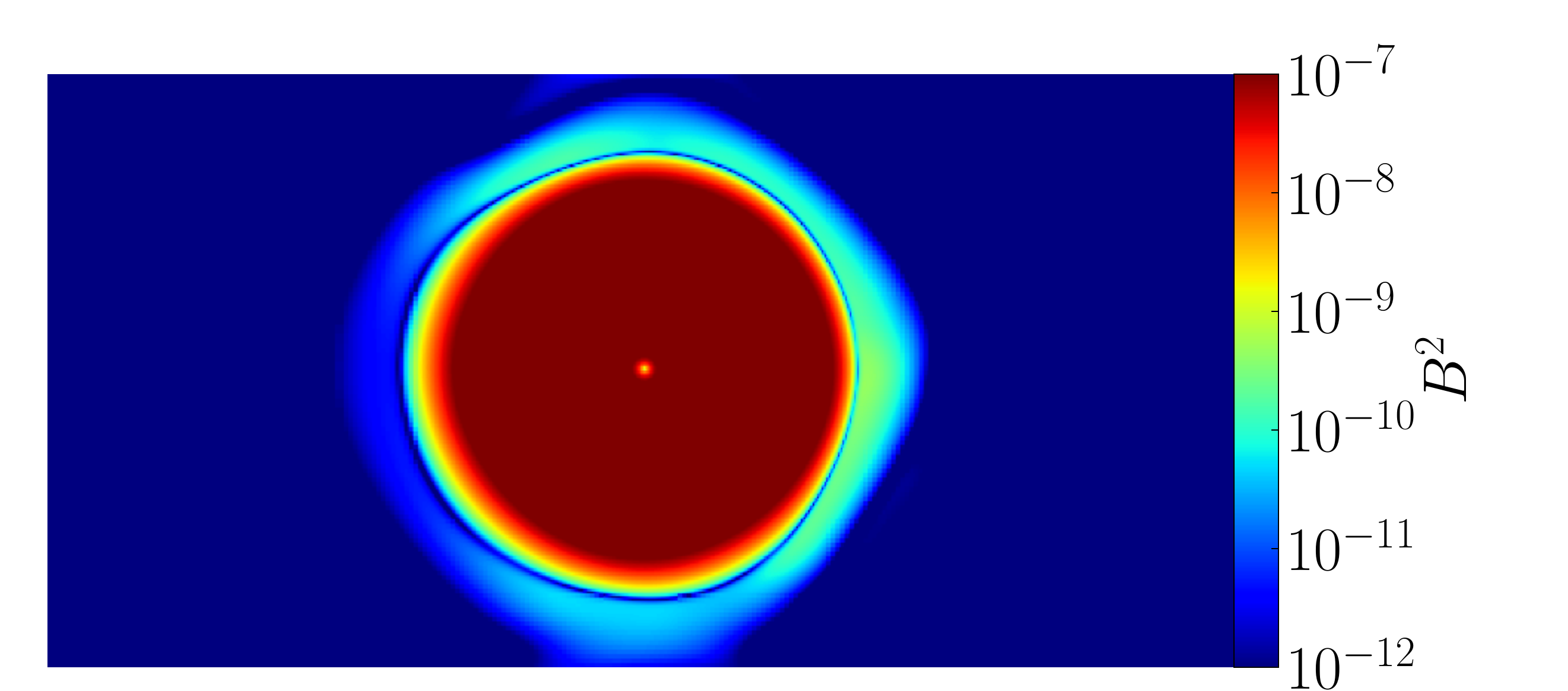} &
    \includegraphics[height=.15\textheight]{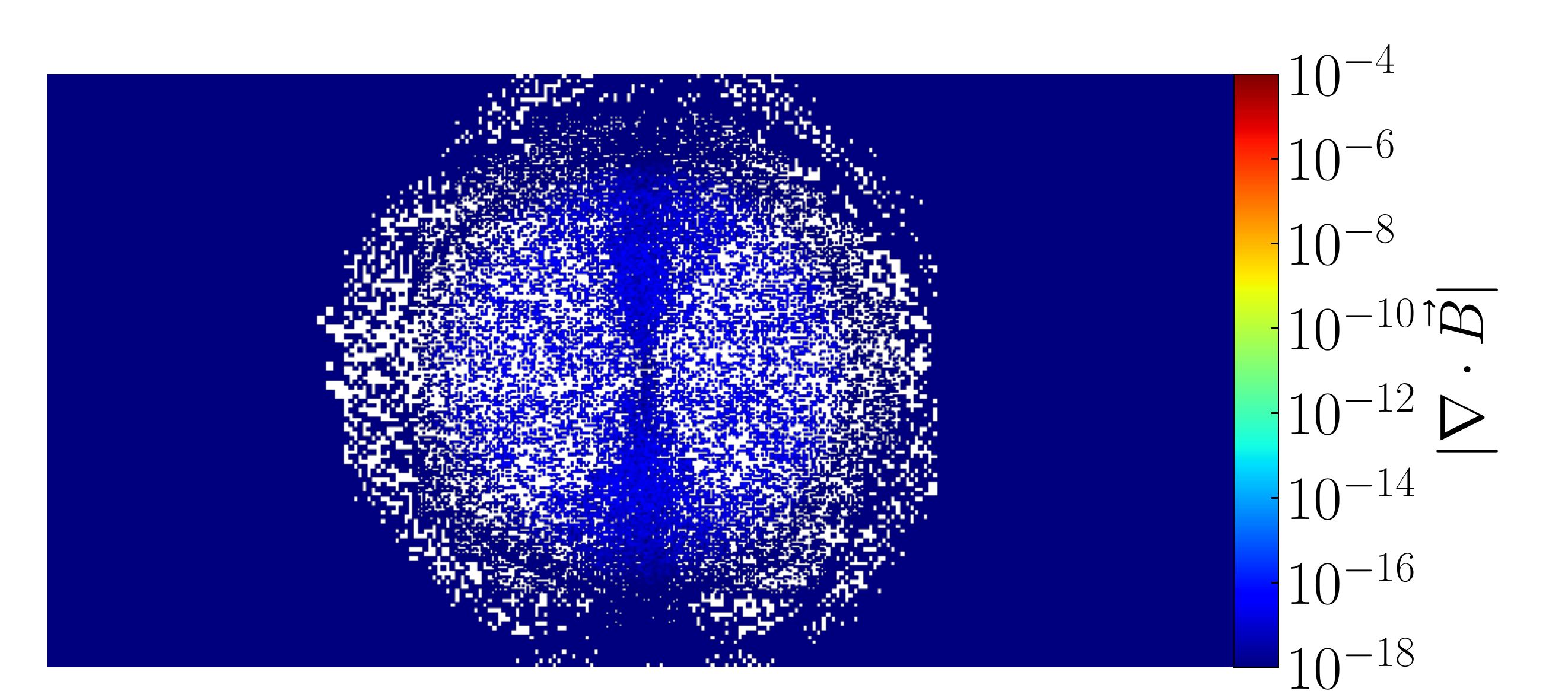} \\
  \end{tabular}
	  \caption{\label{fig:EMHD_loop_advection_2d_compare}
	  Comparison of the square of the magnetic fields $B^2$ (\emph{left column}) and divergence of the magnetic field $||\nabla \cdot \vec{B}||$ (\emph{right column}) for the loop advection test with different divergence-control methods at $t = 100 = 10T$.
	  To compare directly, the range of the corresponding colour bars are set with the ranges $[10^{-7}, 10^{-12}]$ and $[10^{-4}, 10^{-18}]$, for the left and right columns, respectively. 
	  On the right column, the place where $\nabla \cdot \vec{B}$ is identically equal to zero is marked as white. 
	  }
\end{figure*}

\begin{figure}
	\centering
	\includegraphics[width=\columnwidth, angle=0]{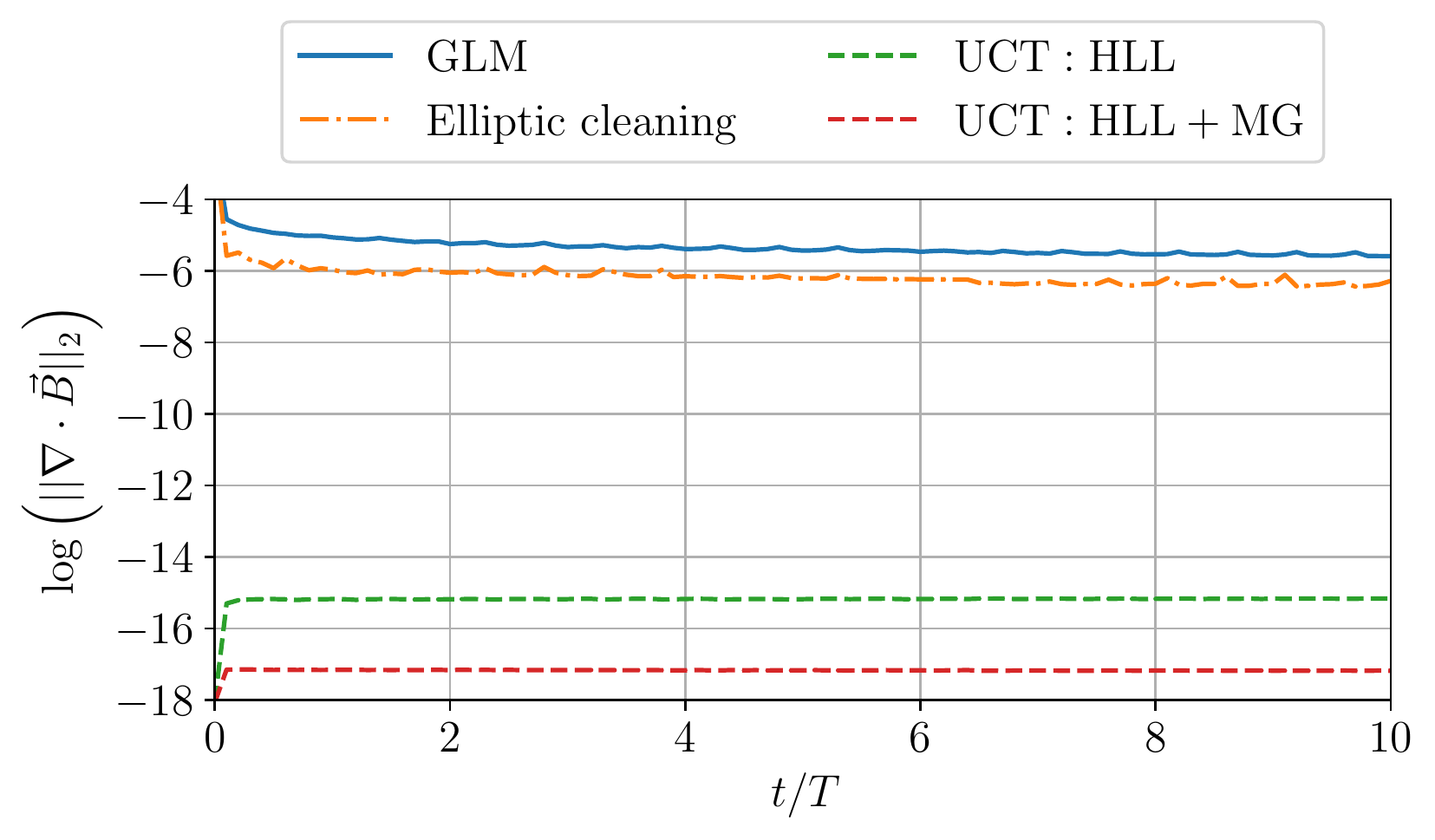}
	\caption{ 
		The $L_2$-norm of $\nabla \cdot \vec{B}$ versus time for the advected field loop test with different divergence-control approaches.
		With the constrained transport scheme, the overall $L_2$-norm of $\nabla \cdot \vec{B}$ is significantly low at the order of $\mathcal{O}\left(10^{-15}\right)$ during the entire simulation.
		Furthermore, combining constrained transport scheme with elliptic cleaning, the magnetic monopoles are further suppressed at the order of $\mathcal{O}\left(10^{-17}\right)$.
		}
	\label{fig:EMHD_loop_advection_2d_L2_divB_compare}	
\end{figure}

Figure~\ref{fig:EMHD_loop_advection_2d_compare} compares the square of the magnetic fields $B^2$ and the absolute value of the divergence of the magnetic field $|\nabla \cdot \vec{B}|$ at the final time ($t = 100 = 10 T$) for the loop advection test with different divergence-control methods.
As shown in figure~\ref{fig:EMHD_loop_advection_2d_compare}, if no divergence-control is activated, the magnetic pressure is worse maintained and ``magnetic monopole'' arises everywhere in the computational domain at the order of $10^{-3}$, which is totally non-physical. 
The evolution of the $L_2$-norm of $\nabla \cdot \vec{B}$ with different divergence-control methods is shown in figure~\ref{fig:EMHD_loop_advection_2d_L2_divB_compare}.
Although there is no significant difference among the results with divergence handling, such huge difference of the divergence of magnetic field could affect the evolution in long term simulations.
Note that the advection test is based on a weakly magnetised loop, the difference between divergence-control methods is expected to be large when the system is strongly magnetised.

\subsection{\label{sec:mag_ns_2d_dynamo}Magnetised neutron star with dynamo effect} 
Here, we present an evolution of a magnetised neutron star with non-vanishing dynamo effect. 
The initial profile used this test, we follow the setup introduced by \cite{2013MNRAS.428...71B}.
In particular, the equilibrium model with a polytropic equation of state with $\Gamma = 2$ and $K=100$ with central rest-mass density $\rho_c = 1.28 \times 10^{-3}$.
The neutron star is non-rotating and is magnetised with magnetic polytropic index $m = 1$ and magnetic coefficient $K_m = 10^{-4}$.
This test problem is simulated with the ideal-gas equation of state with $\Gamma = 2$ and $K=100$ while hyperbolic cleaning is used.
The resistivity is set as $\eta = 0.05$ while the dynamo coefficient is set as $\xi = 0.1$ everywhere in the computational domain.
We simulate this initial model in two-dimensional cylindrical coordinate $(R, z, \varphi)$, where the computational domain covers $0 \leq R \leq 60$ and $-60 \leq z \leq 60$, with the resolution {$N_R \times N_z = 32 \times 64$} and allowing {4 AMR levels} (i.e., an effective resolution of {$256 \times 512 $}), with the final time $t_{\rm{final}} = 200 ~\rm{[code\ unit]}$.
Note that, as pointed out in \cite{2013MNRAS.428...71B}, this setting is for code test only, which is non-physical.

\begin{figure}
	\centering
	\includegraphics[width=\columnwidth, angle=0]{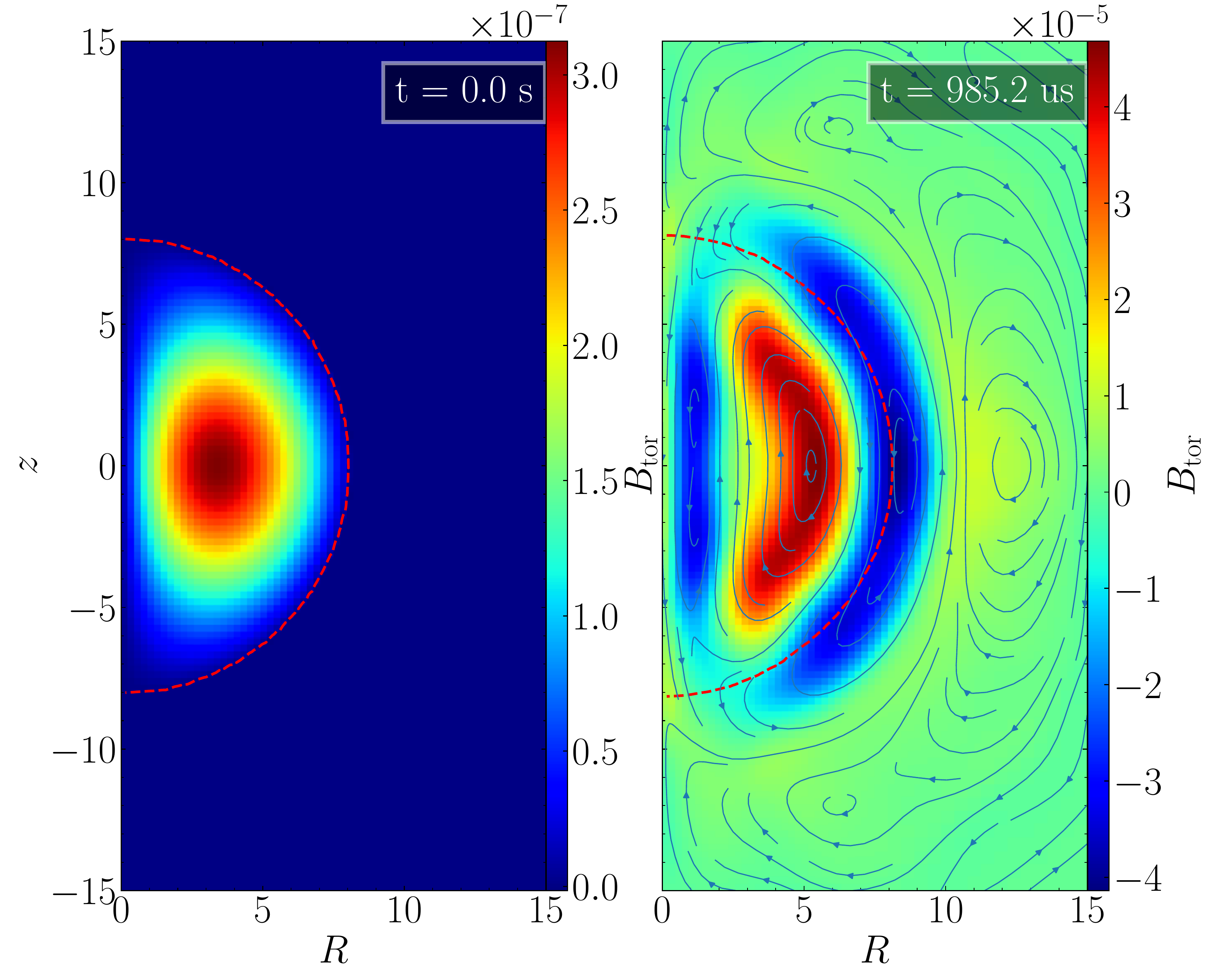}
	\caption{
		Comparison of the initial and final ($t= 200 ~\rm{[code\ unit]} \approx 1 \rm{ms}$) magnetic field.
		The red doted line shows the surface of the star.
		Initially, only toroidal component of magnetic field $\vec{B}_{\rm{tor}}$ exist and no poloidal parts.
		However, the toroidal magnetic field is strongly distorted and grows even in the atmosphere.
		In addition, poloidal magnetic develops (the streamlines), and its energy grows significantly during the evolution.
		These effects are due to the non-physical choice that requires a non-vanishing resistivity and dynamo terms everywhere.
		}
	\label{fig:EMHD_NS_cylindrical_2D_dynamo_time_series}	
\end{figure}

\begin{figure}
	\centering
	\includegraphics[width=\columnwidth, angle=0]{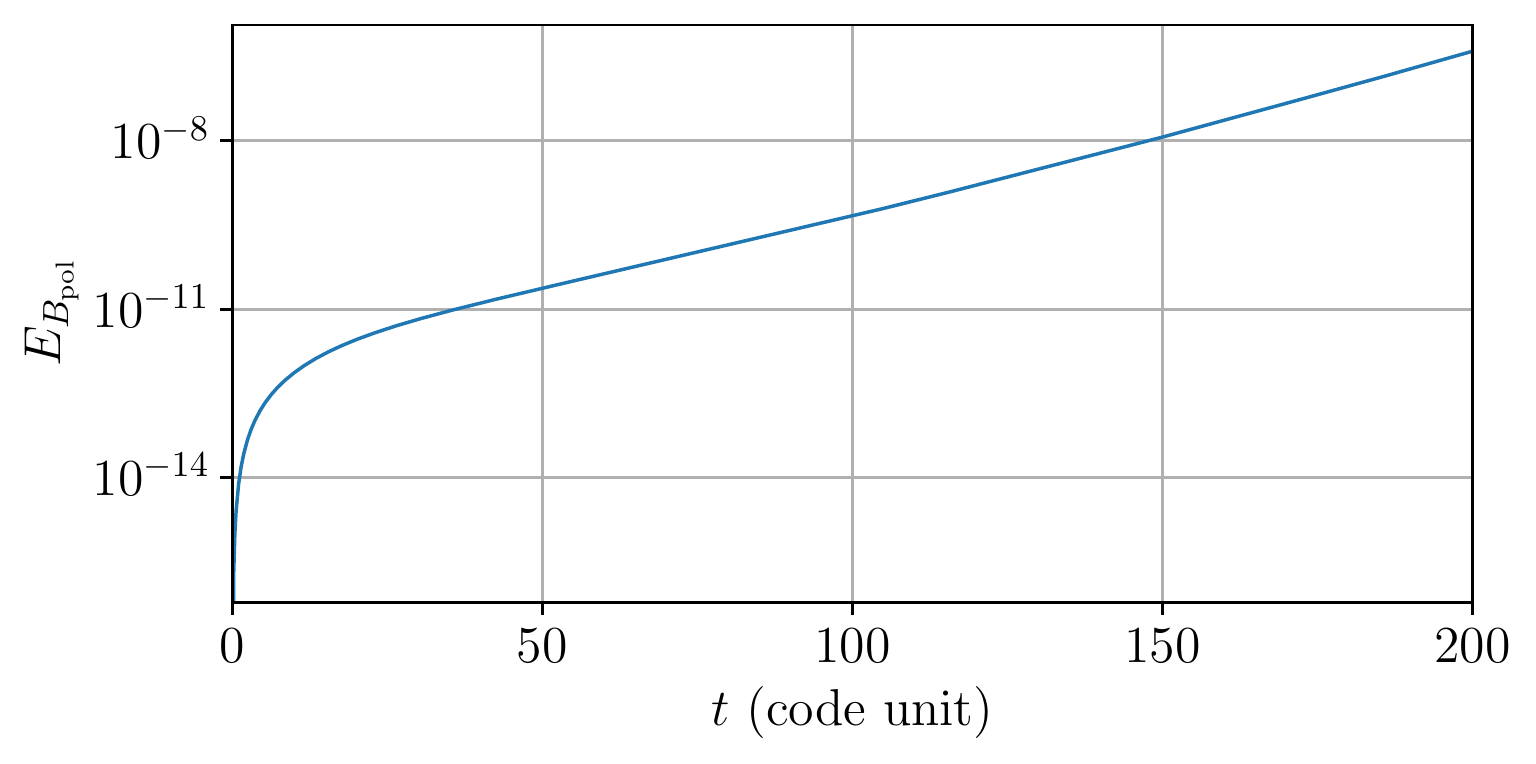}
	\caption{
		The poloidal magnetic develops, and its energy grows significantly during the evolution.
		These effects are due to the non-physical choice that requires a non-vanishing resistivity and dynamo terms everywhere.
		}
	\label{fig:EMHD_NS_cylindrical_2D_dynamo_energy_t}	
\end{figure}

Figure~\ref{fig:EMHD_NS_cylindrical_2D_dynamo_time_series} compares the initial and final ($t= 200 ~\textrm{[code\ unit]} \approx 1 ~\textrm{ms}$) magnetic fields.
Initially, only toroidal component of magnetic field $\vec{B}_{\rm{tor}}$ exist and no poloidal parts.
However, the toroidal magnetic field is strongly distorted and grows even in the atmosphere.
In addition, as shown in figure~\ref{fig:EMHD_NS_cylindrical_2D_dynamo_energy_t}, poloidal magnetic develops, and its energy grows significantly during the evolution.
These effects are due to the non-physical choice that requires a non-vanishing resistivity and dynamo terms everywhere.
Our results agree with \cite{2013MNRAS.428...71B} qualitatively.

\subsection{Magnetised neutron star with purely toroidal field} 
To see the effects due to resistive dissipation, in this section, we present a test similar in the appendix B in \cite{2021PhRvD.103d3022S}.
In particular, the equilibrium model with a polytropic equation of state with $\Gamma = 2$ and $K=110$ with central rest-mass density $\rho_c = 1.387 \times 10^{-3}$.
The neutron star is non-rotating and is magnetised with magnetic polytropic index $m = 1$ and magnetic coefficient $K_m = 1.25 \times 10^{-4}$.
We simulate this initial model in two-dimensional cylindrical coordinate $(R, z, \varphi)$, where the computational domain covers $0 \leq R \leq 120$ and $-120 \leq z \leq 120$, with the resolution {$N_R \times N_z = 64 \times 128$} and allowing {5 AMR levels} (i.e., an effective resolution of {$1024 \times 2048 $}).
This test problem is simulated with the ideal-gas equation of state with $\Gamma = 2$ and $K=100$.

The conductivity $\sigma_{\rm{c}}$ ranges from $10^4$ to $10^7$ everywhere in the computational domain.
Since the Ohmic decay timescale scales linearly with $1/\sigma_{\rm{c}}$, the magnetic energy decreases exponentially with time, which is approximately proportional to $\exp \left(- 2t / \sigma_{\rm{c}} \right)$. 
Note that, as pointed out in \cite{2021PhRvD.103d3022S}, this evolution process is valid only when axial symmetry is assumed.
In general, non-axisymmetric instabilities could occur and significantly affect the evolution of the system.

\begin{figure}
	\centering
	\includegraphics[width=\columnwidth, angle=0]{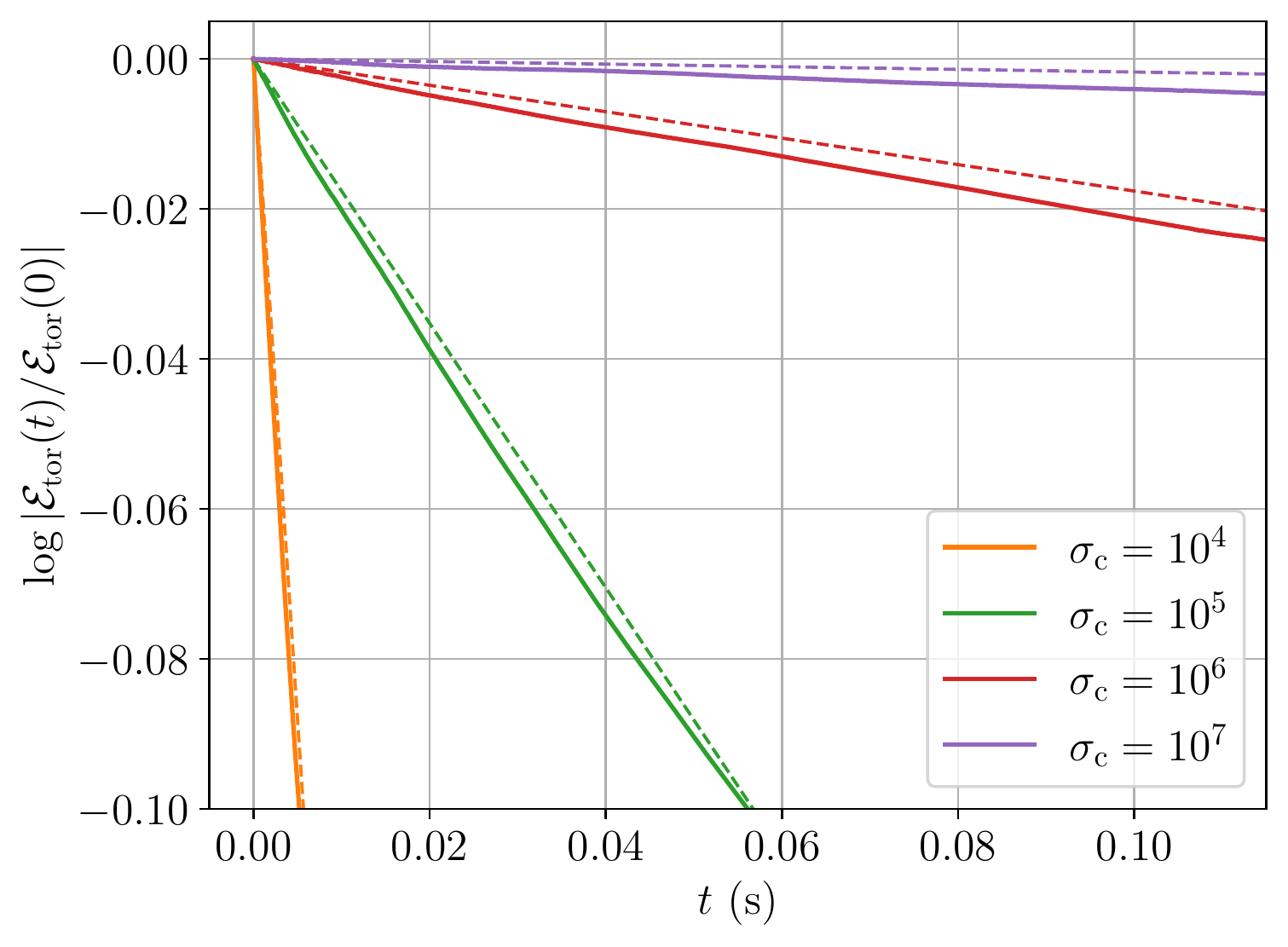}
	\caption{
		The evolution normalised energy of the toroidal field $\mathcal{E}_{\rm{tor}}(t) / \mathcal{E}_{\rm{tor}}(t=0)$ of a magnetised neutron star with a purely toroidal initial field with different conductivity $\sigma_{\rm{c}}$.
		Since the Ohmic decay timescale of magnetic fields scales linearly with $1/\sigma_{\rm{c}}$, the magnetic energy decreases exponentially with time, which is approximately proportional to $\exp \left(- 2t / \sigma_{\rm{c}} \right)$. 
		The magnetic energy decreases exponentially with time (solid lines), the decay rates agree with the expected decay rates (dashed lines). 
		Note that this evolution process is valid only when axial symmetry is assumed.
		}
	\label{fig:EMHD_NS_cylindrical_2D_tor_energy}	
\end{figure}

Figure~\ref{fig:EMHD_NS_cylindrical_2D_tor_energy} shows the evolution normalised energy of toroidal field $\mathcal{E}_{\rm{tor}}(t) / \mathcal{E}_{\rm{tor}}(t=0)$.
The magnetic energy decreases exponentially with time (solid lines), the decay rate matches with the expected decay rate $\exp \left(- 2t / \sigma_{\rm{c}} \right)$ (dashed lines). 

\subsection{\label{sec:mag_ns_2d_resistive}Magnetised neutron star with poloidal field} 
Finally, we present a test of a stable neutron star with extended magnetic field.
The equilibrium model with a polytropic equation of state with $\Gamma = 2$ and $K=100$ with central rest-mass density $\rho_c = 1.28 \times 10^{-3}$.
This neutron star is non-rotating and carry a purely poloidal magnetic field (with $K_{\rm{pol}}= 0.437\times10^{-4}$ in \texttt{XNS}) where the maximum magnetic field is roughly $6.5 \times 10^{13} \rm{G}$.
We simulate this initial model in two-dimensional cylindrical coordinate $(R, z, \varphi)$, where the computational domain covers $0 \leq R \leq 120$ and $-120 \leq z \leq 120$, with the resolution {$N_R \times N_z = 64 \times 128$} and allowing {4 AMR levels} (i.e., an effective resolution of {$512 \times 1024 $}).
This test problem is simulated with the ideal-gas equation of state with $\Gamma = 2$ and $K=100$.

\begin{figure}
	\centering
	\includegraphics[width=\columnwidth, angle=0]{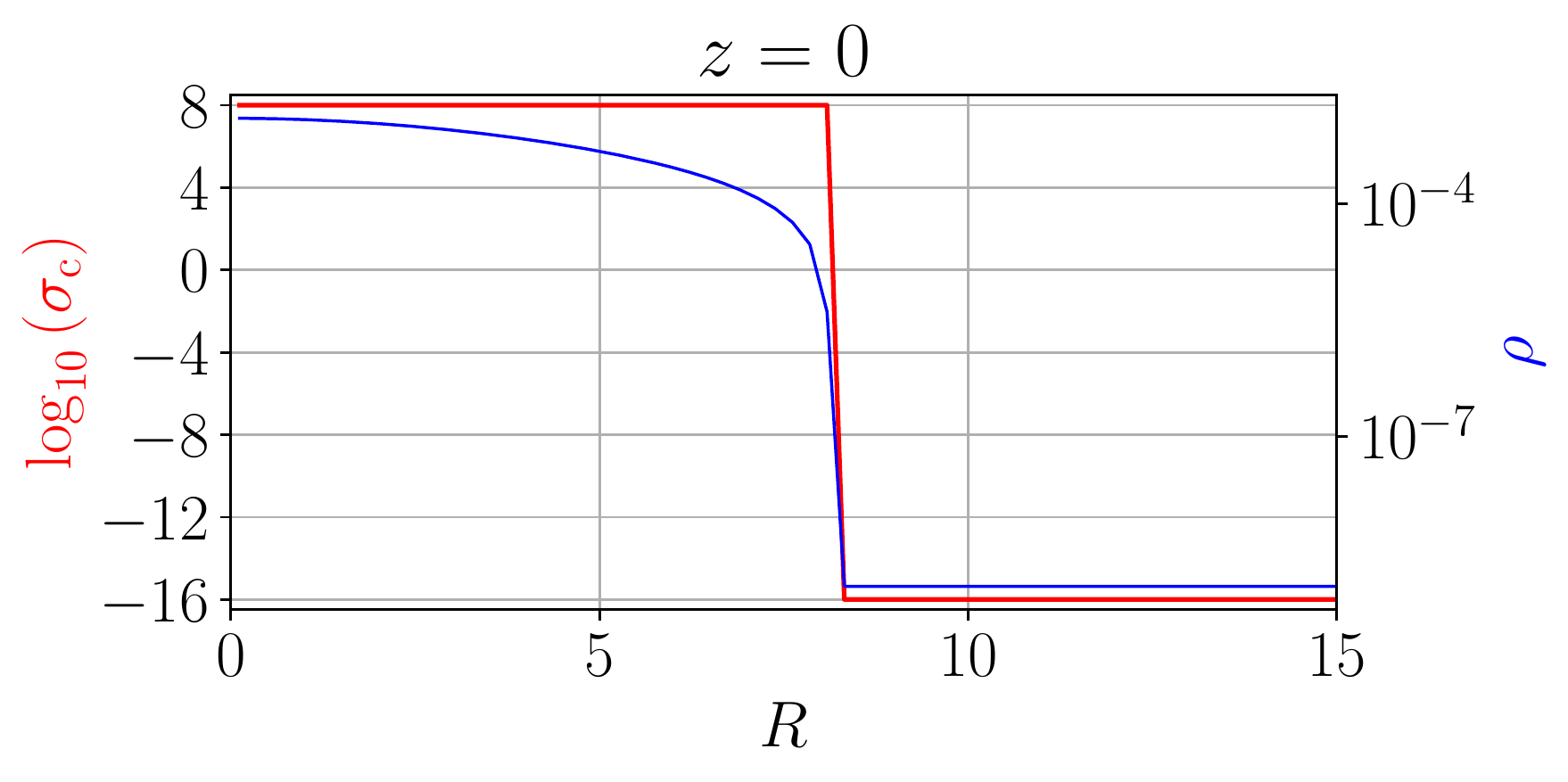}
	\caption{
		An empirical functional form of the conductivity proposed in \cite{2013PhRvD..88d4020D}: $\sigma_{\rm{c}} = \sigma_0 \left( \max\left[ \left( 1 - \frac{D_{\rm{atmo}}}{D} \right), 0\right] \right)^2$, 
		where $D$ here is the conserved density, and the conductivity at the centre of the star is set as $\sigma_0 = 10^8$, which is roughly $2.03 \times 10^{13} \rm{s^{-1}}$ in cgs unit.
		}
	\label{fig:EMHD_NS_cylindrical_2D_pol_sigma}	
\end{figure}

Here, we assume that the conductivity inside the star is extremely high (close to the ideal MHD limit) while the conductivity at the low density regions are expected to be negligibly small.
To model both exterior and interior of the neutron star, the conductivity in different regions needed to be chosen properly.
In this test, we follow the empirical functional form of the conductivity proposed in \cite{2013PhRvD..88d4020D}:
\begin{equation}\label{eq:sigma_D}
	\sigma_{\rm{c}} = \sigma_0 \left( \max\left[ \left( 1 - \frac{D_{\rm{atmo}}}{D} \right), 0\right] \right)^2, 
\end{equation}
where $D$ here is the conserved density, and the conductivity at the centre of the star is set as $\sigma_0 = 10^8$, which is roughly $2.03 \times 10^{13} \rm{s^{-1}}$ in cgs unit.
Note that the Ohmic decay timescale scales linearly with $1/\sigma_{\rm{c}}$, as long as the simulation time is much shorter than the decay timescale, this limit is really closed to ideal-magnetohydrodynamics limit. 
Figure~\ref{fig:EMHD_NS_cylindrical_2D_pol_sigma} shows the conductivity $\sigma_{\rm{c}}$ as a function of density.
Studies of different forms of conductivity function will be explored in the future.
The simulations are carried till $t= 10~\rm{ms}$ with different divergence-free handling approaches, namely, the hyperbolic cleaning, cell centred elliptic cleaning, upwind constrained transport with and without elliptic cleaning.
Since this neutron star is not strongly magnetised, where the Alfv{\'e}n timescale is roughly 26 s (i.e., $\tau_A \sim 2 R \sqrt{2\pi \langle \rho \rangle} / \langle B \rangle \sim 26 \text{s}$), which is much longer than our final simulation time $t=10~\rm{ms}$, no significant changes are expected within this short timescale.

\begin{figure}
	\centering
	\includegraphics[width=\columnwidth, angle=0]{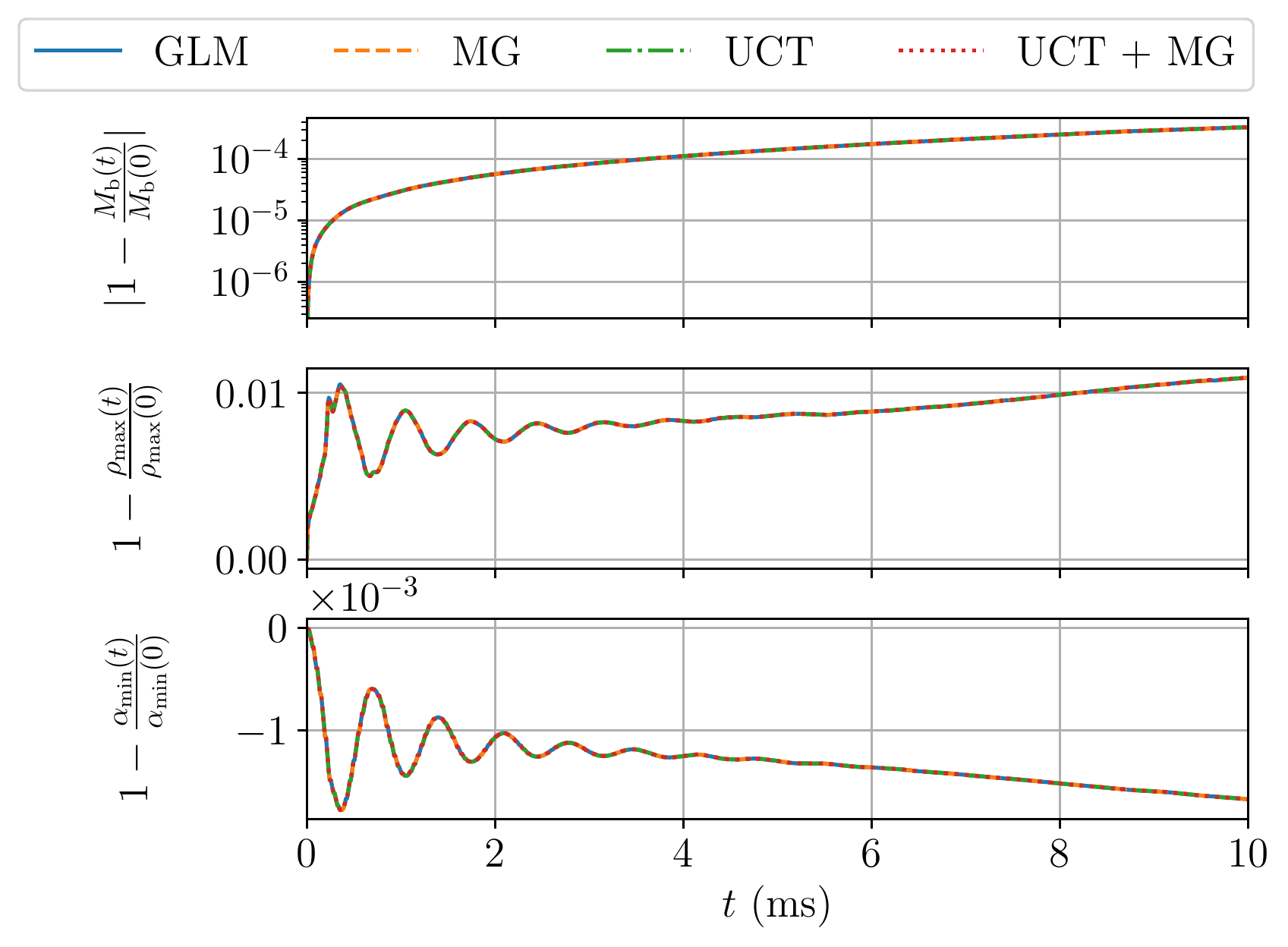}
	\caption{
                \emph{Upper panel}: The relative variation of the rest mass $M_b$ in time.
		\emph{Middle panel}: The relative variation of the maximum density $\rho_{\max}$ in time.
		\emph{Lower panel}: The relative variation of the minimum lapse function $\alpha_{\min}$ in time.
		All the evolution of the fluid are almost identical with all divergence-free handling methods.
		}
	\label{fig:EMHD_NS_cylindrical_2D_pol_combine_hydro}	
\end{figure}

\begin{figure}
	\centering
	\includegraphics[width=\columnwidth, angle=0]{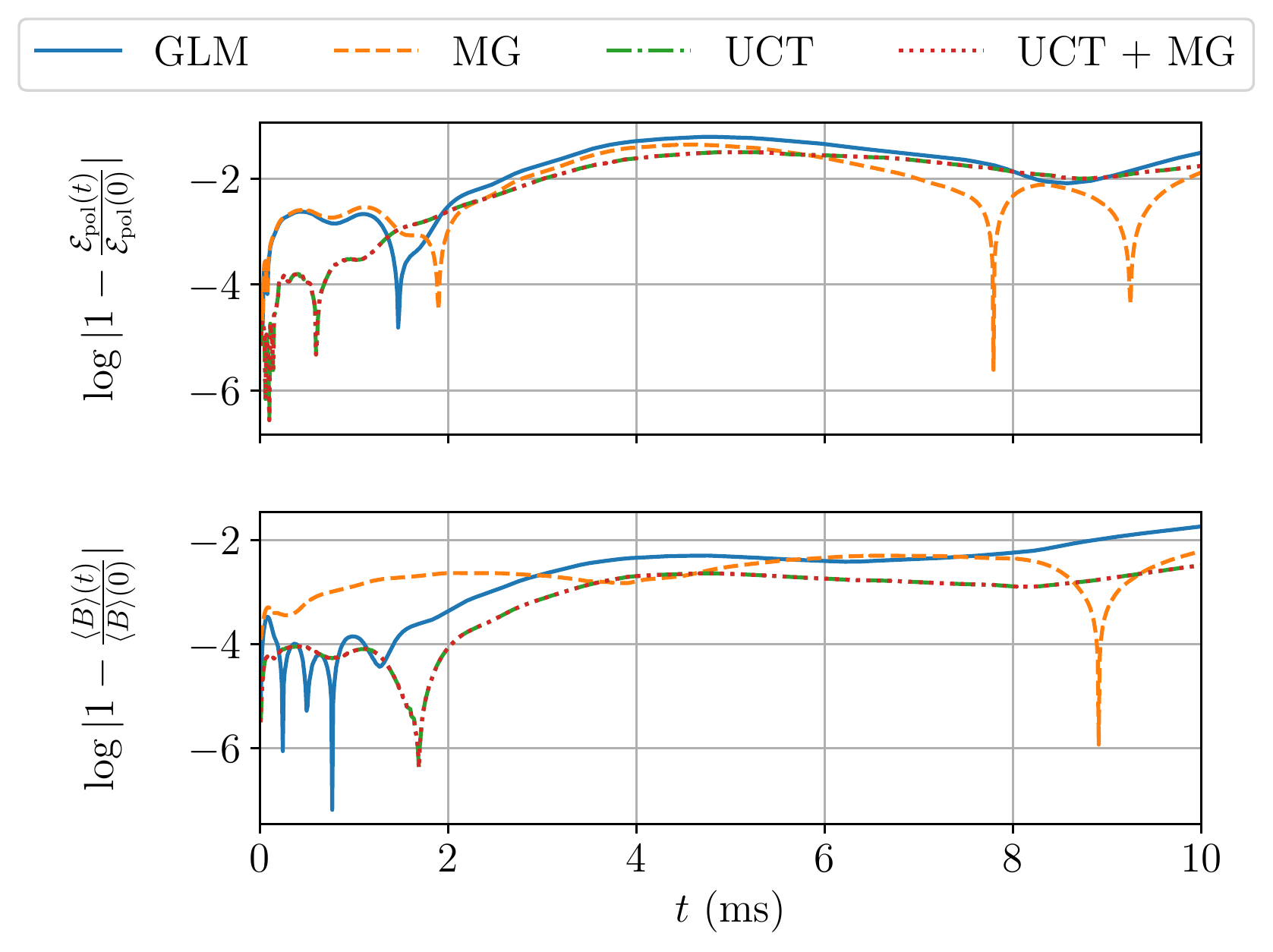}
	\caption{
		\emph{Upper panel}: The relative variation of the magnetic energy of the poloidal field $\mathcal{E}_{\text{pol}}$ in time.
		\emph{Lower panel}: The relative variation of the volume averaged magnetic field $\langle B \rangle$ in time.
		Upwind constrained transport perform slightly better at preserving these quantities.
		Overall, these quantities are preserved well.
		}
	\label{fig:EMHD_NS_cylindrical_2D_pol_combine_mag}	
\end{figure}

\begin{figure*}[h]
	\centering
	\includegraphics[width=\textwidth, angle=0]{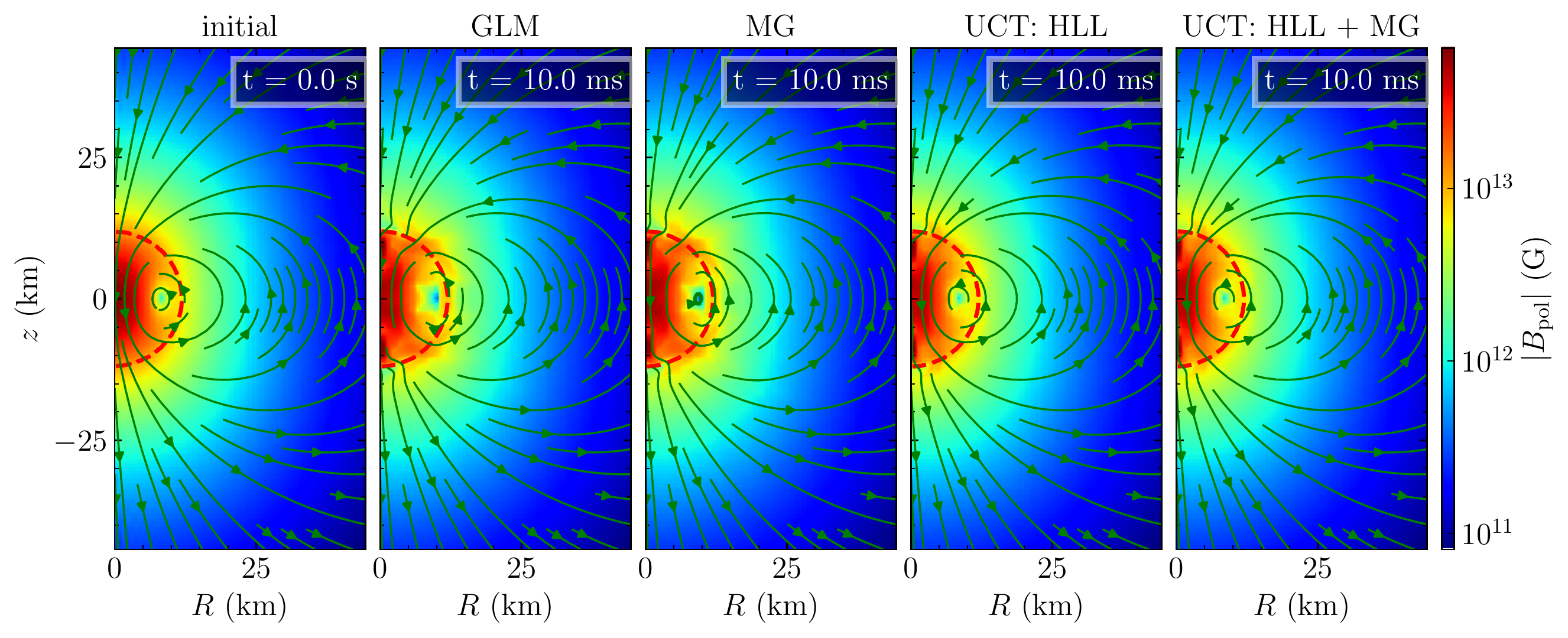}
	\caption{
		Comparison of the initial and final ($t= 10 ~\rm{ms}$) magnetic field with different divergence-free handling approaches, namely, the hyperbolic cleaning, cell centred elliptic cleaning, upwind constrained transport with and without elliptic cleaning.
		The red doted line shows the surface of the star.
		The field is well maintained in all cases except that there are some distortions at the vicinity of the star surface.
		}
	\label{fig:EMHD_NS_cylindrical_2D_pol_time_series}	
\end{figure*}

\begin{figure}
	\centering
	\includegraphics[width=\columnwidth, angle=0]{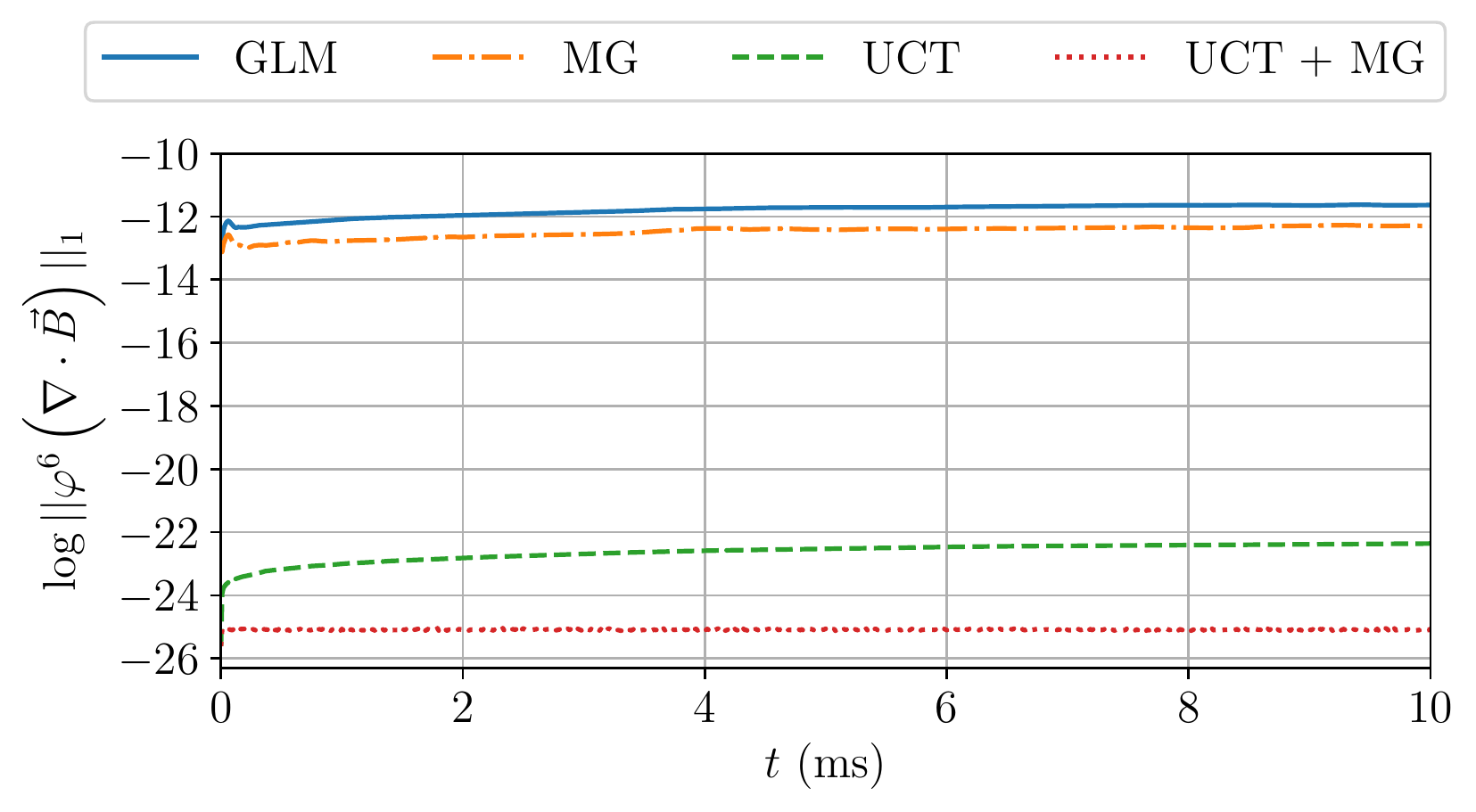}
	\caption{
		Comparison of the evolution of the magnetic monopoles among all divergence-free handling methods we have explored.
		The $L_1$-norms of the magnetic monopoles for GLM is of the order $\mathcal{O}\left( 10^{-12}\right)$, which is roughly one order of magnitude larger than then purely elliptic cleaning case.
		When upwind constrained transport is employed, the $L_1$-norms of the magnetic monopoles are well controlled below $10^{-22}$, which can be significantly suppressed to $\mathcal{O}\left(10^{25}\right)$ (three order of magnitude smaller) if elliptic cleaning is activated on top of upwind constrained transport scheme.
		Although there is no visible difference in the magnetic field with or without elliptic cleaning when upwind constrained transport scheme is used (see figure~\ref{fig:EMHD_NS_cylindrical_2D_pol_time_series}), such huge difference of the magnitude of magnetic monopoles is expected to make difference in long-term simulations.
		}
	\label{fig:EMHD_NS_cylindrical_2D_pol_divB}	
\end{figure}

Figure~\ref{fig:EMHD_NS_cylindrical_2D_pol_combine_hydro} shows the evolution of the total rest mass, maximum density and minimum lapse function in time while figure~\ref{fig:EMHD_NS_cylindrical_2D_pol_combine_mag} shows the evolution of magnetic energy, magnitude of the volume averaged magnetic field.
As shown in figure~\ref{fig:EMHD_NS_cylindrical_2D_pol_combine_hydro}, the dynamics of the fluid are almost identical with all divergence-free handling methods.
In addition, as shown in figure~\ref{fig:EMHD_NS_cylindrical_2D_pol_combine_mag}, upwind constrained transport perform slightly better at preserving the magnetic energy and fields.
Overall, these quantities are preserved well.
Figure~\ref{fig:EMHD_NS_cylindrical_2D_pol_time_series} compares the magnitude and the field lines of the initial and final ($t= 10~\rm{ms}$) magnetic fields.
The field is well maintained in all cases except that there are some distortions at the vicinity of the star surface.
Figure~\ref{fig:EMHD_NS_cylindrical_2D_pol_divB} compares the evolution of the magnetic monopoles among all divergence-free handling methods we have explored.
As shown in figure~\ref{fig:EMHD_NS_cylindrical_2D_pol_divB}, the $L_1$-norms of the magnetic monopoles for GLM is of the order $\mathcal{O}\left( 10^{-12}\right)$, which is roughly one order of magnitude larger than then purely elliptic cleaning case.
When upwind constrained transport is employed, the $L_1$-norms of the magnetic monopoles are well controlled below $10^{-22}$, which can be significantly suppressed to $\mathcal{O}\left(10^{25}\right)$ (three order of magnitude smaller) if elliptic cleaning is activated on top of upwind constrained transport scheme.
Although there is no visible difference in the magnetic field with or without elliptic cleaning when upwind constrained transport scheme is used, such huge difference of the magnitude of magnetic monopoles is expected to make difference in long-term simulations.

\section{Conclusions}\label{sec:conclusions}
We present the implementation of implicit-explicit based general-relativistic resistive magnetohydrodynamics solvers and three divergence-free handling methods available in our code \texttt{Gmunu}.
Three divergence-free handling methods are (i) hyperbolic divergence cleaning through a generalised Lagrange multiplier (GLM); (ii) staggered-meshed constrained transport (CT) schemes and (iii) elliptic cleaning though multigrid (MG) solver which is applicable in both cell-centred and face-centred (stagger grid) magnetic field.

Our implementation has been tested with several benchmarking tests, from special-relativistic to general-relativistic magnetohydrodynamics.
These tests include a magnetic monopole test, a relativistic shock-tube test, the self-similar current sheet, the cylindrical blast wave, the telegraph equation test, the stationary charged vortex test and the steady dynamo test.
In addition, we preform simulations of magnetised neutron stars in dynamical spacetime.
In the magnetised neutron star tests, we demonstrate that different divergence-free handling could significantly affect the evolution of the geometries of the magnetic fields even within a short timescale.
Specifically, the magnetic monopoles are well controlled below double precision when upwind constrained transport is being used.
In addition, the magnetic monopoles can be further suppressed to roughly three order of magnitude smaller by applying elliptic cleaning on top of upwind constrained transport scheme.
In these cases, although there is no visible differences whether elliptic cleaning is activated or not, such huge difference of magnetic monopoles are expected to affect the dynamics of plasmas in long-term simulations.

The general resistive MHD framework enables us to consider a wide range of physical phenomena such as reconnection and dissipation.
More investigations on modelling neutron stars will be presented in the near future.

\begin{acknowledgments}
PCKC thanks Alan Tsz-Lok \textsc{Lam} for helpful discussions. 
This work was partially supported by grants from the Research Grants Council of the Hong Kong (Project No. CUHK14306419), the Croucher Innovation Award from the Croucher Fundation Hong Kong and by the Direct Grant for Research from the Research Committee of the Chinese University of Hong Kong.
\end{acknowledgments}

%



\software{\texttt{Gmunu} \citep{2020CQGra..37n5015C,2021MNRAS.tmp.2378C}  }


\appendix

\section{\label{sec:uct_ideal}The HLL upwind constrained transport in the ideal-MHD cases}
As discussed in \cite{2007A&A...473...11D}, this HLL based upwind constrained transport scheme in ideal-magnetohydrodynamics can be simplified, where only one reconstruction is needed at each time step.
This can be done due to the fact that in the ideal-magnetohydrodynamics cases, the electric field $\hat{E}_i$ can be expressed as
\begin{equation}
	\hat{E}_i = - \epsilon_{ijk} \hat{v}^j B^k = - \eta_{ijk} \sqrt{\hat{\gamma}} \hat{v}^j q_{B^k}.
\end{equation}
In the ideal magnetohydrodynamics module \texttt{grmhd} in \texttt{Gmunu}, we followed the implementation proposed in \cite{2007A&A...473...11D}.
The steps are the following:
\begin{enumerate}[(i)]
	\item Calculate and store the transverse transport velocities at the left and right states (e.g., $\hat{v}^2_{\rm{L}}$, $\hat{v}^2_{\rm{R}}$, $\hat{v}^3_{\rm{L}}$ and $\hat{v}^3_{\rm{R}}$ for the $x^1$-interface), and the characteristic speeds $c^{-}_i$ and $c^{+}_i$, note that here $i$ represent the direction ($i=1,2,3$).
	\item Evaluate a weighted transverse transport velocity, for instance, for the $x^1$-interface,
		\begin{equation}
			\overline{\hat{v}}^j = \frac{c^{-}_i\hat{v}^j_{\rm{R}} + c^{+}_i\hat{v}^j_{\rm{L}}}{c^{-}_i + c^{+}_i}, \qquad \qquad \text{where $i=1$ and $j=2,3$.}
		\end{equation}
	\item Reconstruct the conformally rescaled magnetic fields and weighted transport velocities to the cell edge. 
		This returns $q_{B^{1,2}}^{L,R}$ and $\overline{\hat{v}}^{1,2}_{L,R}$.
	\item Estimate the electric field $\hat{E}_3$ at the cell edge by the formula proposed in \cite{2007A&A...473...11D}:
		\begin{equation}
			\begin{aligned}
				\hat{E}_3 = 
					\sqrt{\hat{\gamma}} \Bigg[
						&- \frac{c^{-}_1\overline{\hat{v}}^1_{\rm{R}} q_{B^2}^{\rm{R}} + c^{+}_1 \overline{\hat{v}}^1_{\rm{L}} q_{B^2}^{\rm{L}} - c^{-}_1 c^{+}_1 \left( q_{B^2}^{\rm{R}} - q_{B^2}^{\rm{L}} \right)}{c^{-}_1 + c^{+}_1} 
					+ \frac{c^{-}_2\overline{\hat{v}}^2_{\rm{R}} q_{B^1}^{\rm{R}} + c^{+}_2\overline{\hat{v}}^2_{\rm{L}} q_{B^1}^{\rm{L}} - c^{-}_2 c^{+}_2 \left( q_{B^1}^{\rm{R}} - q_{B^1}^{\rm{L}} \right)}{c^{-}_2 + c^{+}_2}
				\Bigg].
			\end{aligned}
		\end{equation}
\end{enumerate}


\bibliography{references}{}
\bibliographystyle{aasjournal}



\end{document}